\documentclass[manuscript]{acmart}

\AtBeginDocument{%
  }

\usepackage{graphicx}
\usepackage{dcolumn}
\usepackage{bm}

\usepackage{multirow}
\usepackage{lipsum}
\usepackage{subcaption}
\usepackage{physics}
\usepackage{pgfplots}
\usepackage{pgfplotstable}
\pgfplotsset{compat=1.18}
\usepackage{tikz-cd}
\usepackage{tikz}
 \usepackage{xr}
\usepackage{quantikz}

\usepackage{algorithm}
\usepackage[noend]{algpseudocode}
\usepackage{booktabs}
\usepackage{enumitem}
\setlist{leftmargin=0mm}

\definecolor{avin}{RGB}{230.0, 102.0, 44.0}

\usepackage{soul}

\definecolor{plotgreen}{RGB}{0.0, 116.0, 28.0}
\definecolor{plotred}{RGB}{255.0, 12.0, 20.0}

\begin{document}

\title{Probabilistic Design of Parametrized Quantum Circuits through Local Gate Modifications}

\author{Grier M. Jones}
\orcid{0000-0002-1895-0296}
\authornote{Equal contribution}
\affiliation{%
  \institution{The Edward S. Rogers Sr. Department of Electrical \& Computer Engineering, University of Toronto}
  \streetaddress{10 King's College Rd}
  \city{Toronto}
  \state{ON}
  \postcode{M5S 3G8}
  \country{Canada}
}
\affiliation{%
  \institution{Department of Chemical and Physical Sciences, University of Toronto Mississauga}
  \streetaddress{3359 Mississauga Road}
  \city{Mississauga}
  \state{ON}
  \postcode{L5L 1C6}
  \country{Canada}
}

\author{Aviraj Newatia}
\orcid{0009-0009-6830-0839}
\authornotemark[1]
\affiliation{%
  \institution{The Edward S. Rogers Sr. Department of Electrical \& Computer Engineering, University of Toronto}
  \streetaddress{10 King's College Rd}
  \city{Toronto}
  \state{ON}
  \postcode{M5S 3G8}
  \country{Canada}
}
\affiliation{%
  \institution{Department of Computer Science, University of Toronto}
  \city{Toronto}
  \state{ON}
  \postcode{M5S 2E4}
  \country{Canada}
}
\affiliation{%
  \institution{Vector Institute for Artificial Intelligence}
  \streetaddress{108 College St W1140}
  \city{Toronto}
  \state{ON}
  \postcode{M5G 0C6}
  \country{Canada}
}

\author{Alexander Lao}
\affiliation{%
  \institution{The Edward S. Rogers Sr. Department of Electrical \& Computer Engineering, University of Toronto}
  \streetaddress{10 King's College Rd}
  \city{Toronto}
  \state{ON}
  \postcode{M5S 3G8}
  \country{Canada}
}

\author{Aditya K. Rao}
\orcid{0000-0002-8239-081X}
\affiliation{%
  \institution{The Edward S. Rogers Sr. Department of Electrical \& Computer Engineering, University of Toronto}
  \streetaddress{10 King's College Rd}
  \city{Toronto}
  \state{ON}
  \postcode{M5S 3G8}
  \country{Canada}
}

\author{Viki Kumar Prasad}
\orcid{0000-0003-0982-3129}
\authornote{Corresponding author}
\email{vikikumar.prasad@ucalgary.ca}
\affiliation{%
  \institution{The Edward S. Rogers Sr. Department of Electrical \& Computer Engineering, University of Toronto}
  \streetaddress{10 King's College Rd}
  \city{Toronto}
  \state{ON}
  \postcode{M5G 0C6}
  \country{Canada}
}
\affiliation{%
  \institution{Department of Chemical and Physical Sciences, University of Toronto Mississauga}
  \streetaddress{3359 Mississauga Road}
  \city{Mississauga}
  \state{ON}
  \postcode{L5L 1C6}
  \country{Canada}
}
\affiliation{%
  \institution{Department of Chemistry, University of Calgary}
  \streetaddress{2500 University Drive NW}
  \city{Calgary}
  \state{AB}
  \postcode{T2N 1N4}
  \country{Canada}
}

\author{Hans-Arno Jacobsen}
\orcid{0000-0003-0813-0101}
\authornotemark[2]
\email{arno.jacobsen@utoronto.ca}
\affiliation{%
  \institution{The Edward S. Rogers Sr. Department of Electrical \& Computer Engineering, University of Toronto}
  \streetaddress{10 King's College Rd}
  \city{Toronto}
  \state{ON}
  \postcode{M5S 3G8}
  \country{Canada}
}
\affiliation{%
  \institution{Department of Computer Science, University of Toronto}
  \streetaddress{40 St. George Street}
  \city{Toronto}
  \state{ON}
  \postcode{M5S 2E4}
  \country{Canada}
}

\renewcommand{\shortauthors}{Jones et al.}

\begin{abstract}
Within quantum machine learning, parametrized quantum circuits provide flexible quantum models, but their performance is often highly task‑dependent, making manual circuit design challenging. 
Alternatively, quantum architecture search algorithms have been proposed to automate the discovery of task‑specific parametrized quantum circuits using systematic frameworks.
In this work, we propose an evolution-inspired heuristic quantum architecture search algorithm, which we refer to as the local quantum architecture search.
The goal of the local quantum architecture search algorithm is to optimize parametrized quantum circuit architectures through a local, probabilistic search over a fixed set of gate‑level actions applied to existing circuits. 
We evaluate the local quantum architecture search algorithm on two synthetic function‑fitting regression tasks and two quantum chemistry regression datasets, including the BSE49 dataset of bond separation energies for first‑ and second‑row elements and a dataset of water conformers generated using the data‑driven coupled‑cluster approach. 
Using state‑vector simulation, our results highlight the applicability of local quantum architecture search algorithm for identifying competitive circuit architectures with desirable performance metrics.
Lastly, we analyze the properties of the discovered circuits and demonstrate the deployment of the best‑performing model on state‑of‑the‑art quantum hardware.
\end{abstract}






\maketitle

\section{Introduction}\label{section:Introduction}
In recent years, \textbf{quantum machine learning (QML)} has emerged as a promising subfield at the intersection of artificial intelligence and quantum computing, offering the potential for quantum-enhanced learning across a range of implementations and applications~\cite{schuld2015introduction,Biamonte2017-hp}. 
Previous studies have identified favorable properties of QML algorithms, such as sample-efficiency~\cite{Caro2022-uf, Schuld2021-gb}, quantum mechanical inductive bias~\cite{Bowles2023-cl}, and quantum speedups~\cite{Cao2019-lh,Biamonte2017-hp}. 
\textbf{Variational quantum algorithms (VQAs)} are particularly well suited to near-term noisy intermediate-scale quantum (NISQ) devices due to their comparatively shallow circuit depths and hybrid quantum--classical optimization schemes~\cite{Peruzzo2014-gh,kandala2017hardware,Cerezo2021-fc, McClean2015-nv}. 
Within the broad category of VQAs, \textbf{parametrized quantum circuits (PQCs)} are a promising class of QML models composed of fixed and parametrized gates, which are iteratively optimized using a hybrid quantum--classical scheme~\cite{Benedetti2019-pf}.
PQCs have been shown to produce highly complex behavior at low circuit depths~\cite{Harrow2017-eq,Peruzzo2014-gh} and have demonstrated empirical success in domains such as chemical property and toxicity prediction~\cite{Hatakeyama-Sato2023-lr,Suzuki2020-gv,jin_integrating_2025,bhatia_quantum_2023}.
Despite these advantages, designing task-specific PQC architectures remains a major bottleneck, as performance is highly sensitive to the choice of ansatz and manual design is both time-consuming and error-prone.

To mitigate the challenges of brute-force circuit searches, \textbf{quantum architecture search (QAS)} algorithms have been proposed~\cite{Zhang2022-ho}.
Existing approaches typically rely on global ansatz selection, including those based on reinforcement learning~\cite{Ostaszewski2021-dd, Patel2024-qv}, adversarial networks~\cite{Ma2025-lp}, evolutionary algorithms~\cite{Chivilikhin2020-ea, Lu2021-mx}, among other techniques~\cite{Martyniuk2024-cd}. 
While QAS methods offer automated workflows capable of designing robust, efficient, and practical PQC architectures, the application of such techniques is limited by the trade-off between the number of candidate ans\"{a}tze and the practicality of training, i.e., systems with a greater circuit depth and a larger number of qubits vastly increase the size of this search space~\cite{Du2022-zd, Holmes2022-gc}.
The size of the search space limits global ansatz searches, which are computationally expensive due to the dimensionality of the quantum circuit configuration space.

To address the drawbacks of global ansatz searches, we introduce an iterative, evolution-inspired heuristic quantum architecture search method we call the \textbf{local quantum architecture search} (\textbf{LQAS}). 
Rather than performing global exploration over the full circuit configuration space, LQAS exploits the observation that many expressive and performant ans\"{a}tze can be obtained through localized gate-level modifications of existing circuits. 
LQAS parametrizes such modifications and samples candidate architectures within a local neighborhood, enabling efficient refinement without incurring the cost of global search.
In this study, we evaluate LQAS using state-vector simulation on two function fitting datasets and two quantum chemistry datasets.
The first quantum chemistry dataset is based on the data-driven coupled-cluster (DDCC) method, which utilizes electronic structure data to learn the two-electron excitation amplitudes (or $t_{2}$-amplitudes) of the coupled-cluster singles and doubles (CCSD) wavefunction~\cite{Townsend2019-gx, Jones2023-hz}.
The second dataset, BSE49, is a benchmarking dataset comprised of bond separation energies of 49 different bond types, calculated using the (RO)CBS-QB3 composite method~\cite{Prasad2021-sx}. 
One motivation for exploring quantum chemical datasets is that classical ML approaches have demonstrated aptitude in chemical modeling tasks such as molecular property prediction~\cite{Heid2023-cu, Yang2019-fx, Ramakrishnan2014-oj, Ramakrishnan2015-dg, Hansen2015-en, Unke2019-xz}, drug and catalyst discovery~\cite{Yang2019-jj, Goh2017-ay, Zhong2020-dx, Nandy2021-qg, Jones2023-kh}, and materials design~\cite{Raccuglia2016-tw, Sanchez-Lengeling2018-xt, Butler2018-io} to accelerate computationally expensive tasks~\cite{Janet2020-rl}. 
Additionally, in a recent study,~\citet{Jones2025-kt} highlighted the difficulties of using brute-force searches to find optimal PQCs for the quantum chemical datasets explored herein. 

In our Related Works section (Section \ref{section:relatedworks}), we provide an overview of related QAS methods.
Existing approaches, based on reinforcement learning, evolutionary algorithms, and adversarial optimization, predominantly cast circuit discovery as a global combinatorial optimization problem over discrete and weakly structured architecture spaces.
While such methods can identify expressive ansätze, their scalability is fundamentally constrained by the exponential growth of the search space and by limited reuse of structural information between candidate architectures.
In contrast, the approach proposed in this work is motivated by the observation that high-performing PQCs often lie within locally connected neighborhoods of the architecture space, where incremental gate-level modifications preserve functional structure while enabling targeted refinement.

Lastly, the key contributions of our work are as follows:
\begin{enumerate}
    \item We introduce the LQAS algorithm for designing PQCs via localized gate-level architecture refinement. Since the proposed framework reframes QAS as an iterative local navigation problem rather than a global discovery task, LQAS offers a more scalable and hardware-relevant alternative to existing global search strategies.
    \item We perform an extensive evaluation using state-vector simulation, including benchmarking LQAS on function fitting tasks and two quantum chemistry datasets. Both quantum chemistry datasets offer unique insights into the performance of PQCs using domain-specific applications.
    \item We present real-device experiments on IBM quantum processors that, while exhibiting degraded performance relative to state-vector simulation, provide a valuable baseline for assessing the impact of hardware noise and for benchmarking future algorithmic and hardware improvements.
\end{enumerate}


\section{Background}\label{section:Background}
A common starting point for QML tasks is selecting an available template ansatz that is low-depth and suitable for general learning tasks. 
In this work, we adopt the hardware-efficient ansatz (HEA) of~\citet{kandala2017hardware} as the baseline to improve using our LQAS algorithm.
The HEA circuits used throughout our study consist of three parts: a single-angle feature encoding layer, a parametrized layer that is iteratively updated, and a cyclic sequence of two-qubit entanglement gates. 
For an $n$-qubit system, with qubits labeled $q_{0}, \ldots, q_{n-1}$, the feature matrix, $\mathbf{X}$, has dimensions of $dim(\mathbf{X})=N_{\mathrm{samples}} \times n$, where $N_{\mathrm{samples}}$ denotes the number of samples (or rows) in $\mathbf{X}$ and the number of the features (columns) corresponds to the number of qubits.
The feature matrix is mapped onto a quantum device using a single-angle encoding layer, denoted as,
\begin{equation}
    \bigotimes_{j=0}^{n-1} RX_{j}(x_{ij}),
    \label{eq:AngleEncoding}
\end{equation}
where $i=0,\ldots,N_{\mathrm{samples}}-1$.
The second component of the circuit is composed of parametrized rotation gates that are iteratively updated throughout the optimization process, defined as,
 \begin{equation}
    \bigotimes_{i=0}^{n-1} RY_{i}(\theta_{i0}) RZ_{i}(\theta_{i1}) RY_{i}(\theta_{i2}),
    \label{eq:RotationLayer}
\end{equation}
where $\theta_{i0\cdots2}$ are optimizable parameters.
The last layer of our HEA circuit is the entanglement layer, also known as the dynamic layer, which is iteratively updated throughout the LQAS algorithm. 
Since we use a cyclic pattern of $CNOT$ gates as our entanglement layer, we represent this as $CNOT_{ij}$, where $i=0,\ldots,n-1$ and
\begin{equation}
    j = 
    \left\{
    \begin{array}{cc}
    i+1 & i<n-1 \\
    0  & i=n-1   \\
    \end{array} 
    \right. 
\end{equation}

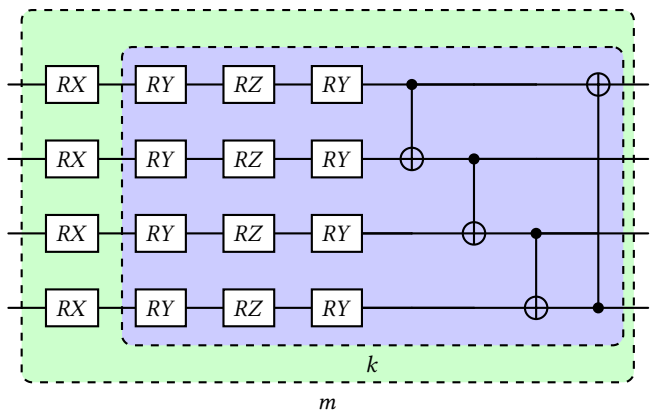
\begin{figure}[H]
    \centering
    
    \begin{quantikz}
        & \gate{RX}\gategroup[4,steps=8,style={dashed,rounded corners,fill=green!20, outer sep=6pt,inner xsep=6pt,
        inner ysep=18pt},background,label style={label position=below,anchor=north,yshift=-0.1cm}]{{\sc $m$}} & \gate{RY}\gategroup[4,steps=7,style={dashed,rounded corners,fill=blue!20, inner xsep=2pt},background,label style={label position=below,anchor=north,yshift=-0.2cm}]{{\sc $k$}} & \gate{RZ} & \gate{RY} & \ctrl{1} & \qw      & \qw      & \targ{3} & \\
        & \gate{RX} & \gate{RY} & \gate{RZ} & \gate{RY} & \targ{}  & \ctrl{1} & \qw      & \qw & \\
        & \gate{RX} & \gate{RY} & \gate{RZ} & \gate{RY} & \qw      & \targ{}  & \ctrl{1} & \qw & \\
        & \gate{RX} & \gate{RY} & \gate{RZ} & \gate{RY} & \qw      & \qw      & \targ{}  & \ctrl{-3} & 
    \end{quantikz}
    
    \caption{An example of a four-qubit hardware-efficient circuit. The variational layer (purple) consists of parameterized rotation and entangling gates and can be repeated $k$-times. When combined with a data-encoding layer, it forms a data re-uploading layer, which can be repeated $m$-times.}
    \label{fig:HEA}
\end{figure}

To introduce additional parameters, nonlinearity, and expressibility of the circuits, we explore two possible expansion strategies. 
The first expansion strategy combines the parametrized rotation gates (Eq. \ref{eq:RotationLayer}) with the entanglement layer to form a variational layer that can be repeated $k$-times to introduce more parameters into the circuit.
The second strategy, which is equivalent to the well-known Universal Approximation Theorem for classical neural networks~\cite{perez2020data}, is data re-encoding (also referred to as data re-upload).
The goal of data re-encoding is to introduce additional nonlinearity in the model by repeating the feature encoding layer (Eq. \ref{eq:AngleEncoding}) and variational layer $m$-times.
Both expansion strategies can be unified mathematically as,
\begin{equation}
    \prod_{m} \left( \bigotimes_{j=0}^{n-1} RX_{j}(x_{ij}) \prod_{k} \left( \bigotimes_{i=0}^{n-1} RY_{i}(\theta_{i0}) RZ_{i}(\theta_{i1}) RY_{i}(\theta_{i2}) CNOT_{ij} \right) \right),
\end{equation}
which we refer to as HEA-$k$-$m$.
An example of a four-qubit HEA-$k$-$m$ circuit is highlighted in Figure~\ref{fig:HEA}.

For regression-based tasks, like those targeted in this study, the types of functions that may be fitted depend on the capability of the circuit structure, implemented with quantum gate operations, to represent the true Fourier series~\cite{Schuld2021-sf}. 
Thus, minor modifications of the ansatz gate operations may lead to a significant change in the expressibility and optimization of the circuit. 
Therefore, this work proposes a systematic method to sample candidates from a given base ansatz to select a circuit suited towards a learning task.

\section{Related Works}\label{section:relatedworks}
The task of designing optimal quantum circuits using automated methodologies is a decades-long pursuit, dating back to the late 1990s and early 2000s, and includes QAS strategies based on reinforcement learning and evolutionary algorithms,  among other approaches that are less directly related to our proposed method.
In reinforcement learning-based approaches, QAS is a Markov decision process modeled as a sequence of gate-selection decisions~\cite{Martyniuk2024-cd}. 
An approach proposed by~\citet{Ostaszewski2021-dd} shapes the search space by formulating the episodic reward dependent on the circuit energy and the exploration space as a two-dimensional matrix, iteratively adding single gates to the PQC. 
Alternatively, the curriculum-based reinforcement learning QAS (CRLQAS)~\cite{Patel2024-qv} approach uses a 3D circuit-encoding architecture to guide search-space exploration, producing shorter final circuits and regular stochastic perturbations to speed up policy convergence.

Another class of algorithms, more closely related to our proposed method, are evolutionary algorithms. 
Evolutionary algorithms, particularly genetic algorithms are a popular approach for QAS inspired by biological evolution, such as crossover and mutation. 
For example, the multiobjective genetic variational quantum eigensolver (MoG-VQE)~\cite{Chivilikhin2020-ea} approach explores PQC topology optimization using the multi-objective non-dominated sorting genetic algorithm (NSGA-II)~\cite{Deb2002-zx}, which directly attempts to find the Pareto optimal selections to the specified objectives. 
While MoG-VQE successfully minimizes the two-qubit gate count in PQCs, the authors raise concerns about the optimal choice of gate blocks and parametrization of the circuits.
Similarly, other works automate the design of quantum feature maps using quantum support vector machines using the NSGA-II algorithm~\cite{Altares-Lopez2021-yc} based on a different formulation of multi-objective Pareto optimization. 
The Markovian quantum neuroevolution (MQNE) algorithm~\cite{Lu2021-mx} is based on the neuroevolution of augmenting topologies (NEAT) algorithm~\cite{Stanley2002-rj}, and maps quantum circuits to directed paths in a graph.
MQNE restricts searches on the graph based on the parameters of the prescribed modeling task and uses an evolutionary algorithm to evolve its path selection. 
Other works have also explored the robustness of discovered quantum circuits to noise~\cite{Wang2022-hi} and resource-efficiency~\cite{Huang2022-as}.
Existing works in QAS approach the search space size problem by defining smaller ans\"{a}tze search spaces using a supernet, a method from neural architecture search, to produce an ans\"{a}tze pool~\cite{Du2022-zd}.
In practice, this approach will yield a sparse sample of candidate ans\"{a}tze due to the size of the search space. 
This work marks remarkable headway in QAS, but the training of a supernet, regardless of the ansatz weight sharing structure, lessens the practicality of this approach in larger qubit settings due to the intractable amount of training data required.

Existing QAS methods predominantly cast circuit discovery as a global combinatorial optimization problem over discrete and weakly structured architecture spaces.
Approaches based on reinforcement learning, evolutionary algorithms, and adversarial optimization, therefore attempt to explore or approximate Pareto-optimal solutions across large candidate sets of PQCs, incurring substantial computational cost as circuit depth and system size increase.
While such methods can identify expressive ans\"{a}tze, their scalability is fundamentally constrained by the exponential growth of the search space and by limited reuse of structural information between candidate architectures.
In contrast, the approach proposed in this work is motivated by the observation that high-performing PQCs often lie within locally connected neighborhoods of the architecture space, where incremental gate-level modifications preserve functional structure while enabling targeted refinement. 
By explicitly exploiting this locality, the proposed LQAS framework reframes QAS as an iterative local navigation problem rather than a global discovery task, offering a more scalable and hardware-relevant alternative to existing global search strategies.

\section{Methods}\label{section:methods}
\subsection{Localized Quantum Architecture Search (LQAS)}\label{subsection:LQAS}
The LQAS algorithm is an evolution-inspired, heuristic algorithm based on a local, ansatz-level architecture search. 
LQAS designs task-specific PQCs by modifying the sequence of quantum gate operations representing the template ansatz.
As an evolution-inspired algorithm, LQAS starts with a base population (``template'') which is then modified and evaluated to produce the next ``generation'' of candidate ans\"{a}tze.
LQAS terminates after a user-defined number of iterations, using the top-performing ans\"{a}tze from the previous generation as the base population.

\begin{figure*}[t!]
    \centering
    \resizebox{\textwidth}{!}{
    \begin{quantikz}
        \lstick{0}& \gate{RX} & \gate{RY} & \gate{RZ} & \gate{RY} & \ctrl{1} & \qw      & \qw      & \targ{3} & \gate{RX} & \gate{RY} & \gate{RZ} & \ctrl{1} & \qw      & \qw      & \targ{3} & \gate{RX} & \gate{RY} & \gate{RZ} & \gate{RY} & \ctrl{1} & \qw      & \qw      & \targ{3} & \gate{RX} & \gate{RY} & \gate{RZ} & \ctrl{1} & \qw      & \qw      & \targ{3} & \gate{RX} & \gate{RY} & \gate{RZ} & \gate{RY} & \ctrl{1} & \qw      & \qw      & \targ{3} & \gate{RX} & \gate{RY} & \gate{RZ} & \ctrl{1} & \qw      & \qw      & \targ{3} & \meter{} \\
        \lstick{1}& \gate{RX} & \gate{RY} & \gate{RZ} & \gate{RY} & \targ{}  & \ctrl{1} & \qw      & \qw      & \gate{RX} & \gate{RY} & \gate{RZ} & \targ{}  & \ctrl{1} & \qw      & \qw      & \gate{RX} & \gate{RY} & \gate{RZ} & \gate{RY} & \targ{}  & \ctrl{1} & \qw      & \qw      & \gate{RX} & \gate{RY} & \gate{RZ} & \targ{}  & \ctrl{1} & \qw      & \qw      & \gate{RX} & \gate{RY} & \gate{RZ} & \gate{RY} & \targ{}  & \ctrl{1} & \qw      & \qw      & \gate{RX} & \gate{RY} & \gate{RZ} & \targ{}  & \ctrl{1} & \qw      & \qw      & \qw \\
        \lstick{2}& \gate{RX} & \gate{RY} & \gate{RZ} & \gate{RY} & \qw      & \targ{}  & \ctrl{1} & \qw      & \gate{RX} & \gate{RY} & \gate{RZ} & \qw      & \targ{}  & \ctrl{1} & \qw      & \gate{RX} & \gate{RY} & \gate{RZ} & \gate{RY} & \qw      & \targ{}  & \ctrl{1} & \qw      & \gate{RX} & \gate{RY} & \gate{RZ} & \qw      & \targ{}  & \ctrl{1} & \qw      & \gate{RX} & \gate{RY} & \gate{RZ} & \gate{RY} & \qw      & \targ{}  & \ctrl{1} & \qw      & \gate{RX} & \gate{RY} & \gate{RZ} & \qw      & \targ{}  & \ctrl{1} & \qw      & \qw \\
        \lstick{3}& \gate{RX} & \gate{RY} & \gate{RZ} & \gate{RY} & \qw      & \qw      & \targ{}  & \ctrl{-3} & \gate{RX} & \gate{RY} & \gate{RZ} & \qw      & \qw      & \targ{}  & \ctrl{-3} & \gate{RX} & \gate{RY} & \gate{RZ} & \gate{RY} & \qw      & \qw      & \targ{}  & \ctrl{-3} & \gate{RX} & \gate{RY} & \gate{RZ} & \qw      & \qw      & \targ{}  & \ctrl{-3} & \gate{RX} & \gate{RY} & \gate{RZ} & \gate{RY} & \qw      & \qw      & \targ{}  & \ctrl{-3} & \gate{RX} & \gate{RY} & \gate{RZ} & \qw      & \qw      & \targ{}  & \ctrl{-3} & \qw
    \end{quantikz}

    }
    
    \vspace{10pt}
    
    \resizebox{\textwidth}{!}{
    \begin{quantikz}
        \lstick{0}& \gate{RX} & \gate{RY} & \gate{RZ} & \gate{RY} & \ctrl{1} & \qw      & \qw & \qw     & \targ{3} & \gate{RX} & \gate{RY} & \gate{RZ} & \ctrl{1} & \qw      & \qw      & \targ{3} & \gate{RX} & \gate{RY} & \gate{RZ} & \gate{RY} & \ctrl{1} & \gate{RZ}     & \qw      & \targ{3} & \gate{RX} & \gate{RY} & \gate{RZ} & \ctrl{1} & \qw      & \qw      & \gate{RY} & \gate{RX} & \gate{RY} & \gate{RZ} & \gate{RY} & \ctrl{1} & \qw & \ctrl{1} & \qw    & \qw      & \targ{3} & \gate{RX} & \gate{RY} & \gate{RZ} & \ctrl{1} & \qw      & \qw      & \targ{3} & \meter{} \\
        \lstick{1}& \gate{RX} & \gate{RY} & \gate{RZ} & \gate{RY} & \targ{}  & \ctrl{1} & \gate{RX} & \qw      & \qw      & \gate{RX} & \gate{RY} & \gate{RZ} & \targ{}  & \ctrl{1} & \qw      & \qw      & \gate{RX} & \gate{RY} & \gate{RZ} & \gate{RY} & \targ{}  & \ctrl{1} & \qw      & \qw      & \gate{RX} & \gate{RY} & \gate{RZ} & \targ{}  & \ctrl{1} & \qw      & \qw      & \gate{RX} & \gate{RY} & \gate{RZ} & \gate{RY} & \targ{}  & \ctrl{1} & \targ{} & \targ{} & \qw      & \qw      & \gate{RX} & \gate{RY} & \gate{RZ} & \gate{RZ}  & \ctrl{1} & \qw      & \qw      & \qw \\
        \lstick{2}& \gate{RX} & \gate{RY} & \gate{RZ} & \gate{RY} & \qw      & \targ{}  & \ctrl{1} & \qw & \qw     & \gate{RX} & \gate{RY} & \gate{RZ} & \qw      & \gate{RY}  & \ctrl{1} & \qw      & \gate{RX} & \gate{RY} & \gate{RZ} & \gate{RY} & \qw      & \targ{}  & \ctrl{1} & \qw      & \gate{RX} & \gate{RY} & \gate{RZ} & \qw      & \targ{}  & \ctrl{1} & \qw      & \gate{RX} & \gate{RY} & \gate{RZ} & \gate{RY} & \qw      & \targ{} & \qw & \ctrl{-1}   & \ctrl{1} & \qw      & \gate{RX} & \gate{RY} & \gate{RZ} & \qw      & \targ{}  & \ctrl{1} & \qw      & \qw \\
        \lstick{3}& \gate{RX} & \gate{RY} & \gate{RZ} & \gate{RY} & \qw      & \qw      & \targ{}  & \gate{RZ} & \ctrl{-3} & \gate{RX} & \gate{RY} & \gate{RZ} & \qw      & \qw      & \gate{RX}  & \ctrl{-3} & \gate{RX} & \gate{RY} & \gate{RZ} & \gate{RY} & \gate{RY}      & \qw      & \targ{}  & \ctrl{-3} & \gate{RX} & \gate{RY} & \gate{RZ} & \qw      & \qw      & \targ{}  & \ctrl{-3} & \gate{RX} & \gate{RY} & \gate{RZ} & \gate{RY} & \qw      & \qw & \qw & \qw     & \targ{}  & \ctrl{-3} & \gate{RX} & \gate{RY} & \gate{RZ} & \qw      & \qw      & \targ{}  & \ctrl{-3} & \qw
    \end{quantikz}
    }
    
    \caption{This figure represents the kinds of circuit-level modifications made by the sampling procedure of LQAS. The top circuit represents a member of the base population on an arbitrary iteration of LQAS. This circuit is an HEA-$2$ ansatz with 3 time depth re-encoding (HEA-2-3). The bottom circuit is a sampled ansatz that maintains the general gate configuration of the top circuit with modifications sampled ($p_\text{add} = 0.2$, $p_\text{remove} = 0.02$, $p_\text{switch}=0.1$, $p_\text{move} = 0.01$) through LQAS's ansatz modification procedure.}
    \label{fig:ansatz-modification}
\end{figure*}
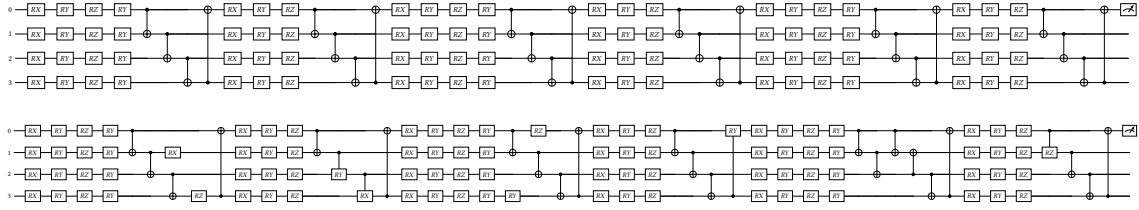

In LQAS, the search space is defined by a set of modification actions, each parametrized with an associated probability, defined as $\mathbf{p_\text{add}}$, $\mathbf{p_\text{remove}}$, $\mathbf{p_\text{switch}}$, and $\mathbf{p_\text{move}}$.
On each iteration, LQAS randomly samples a selection of circuits modified from a baseline circuit using the aforementioned actions and trains the circuits on the specified task.
The probabilities of the modification actions define the sampling neighborhood of the base population.
Upon completion of training, each circuit is assessed based on its performance across a set of desirable metrics, such as validation set loss.
LQAS then chooses the top-$m$ circuits by performance ($m$ is a hyperparameter) as the baseline population for the next iteration of the algorithm.
A new set of modified circuits is then produced by sampling circuit modification actions on the new baseline population.
An example of this modification procedure is highlighted in Figure~\ref{fig:ansatz-modification} using HEA-2 with 3 time depth re-encoding (HEA-2-3) based on~\citet{Schuld2021-sf}.
The iterative sampling and training process of LQAS allows the algorithm to refine the baseline ansatz, building on optimally behaving modification actions on each iteration.

The baseline ansatz can be denoted mathematically as $A_0$, which is input into the sampling process and is assumed to be a previously designed PQC adapted from previous work (such as HEA).
The aim is to improve the performance of $A_0$ according to some performance metric through random modifications of some or all of the gates of $A_0$.
To represent the modifiable gates of $A_0$, we arrange them in temporal order as a sequence (for $t=1,\ldots,n$), written as:
\begin{equation}
    A_0 = (g_1, \ldots, g_n).
\end{equation}
 We define a set of four gate actions, each with random variables parametrized by probabilities that are hyperparameters of LQAS.
 The first action is gate addition, denoted as $\mathbf{p_1 = p_\text{add}}$.
 If $g_t$ acts on $m$ qubits, the action randomly chooses an $m$ qubit gate to insert after $g_t$ (e.g., given a CNOT gate $g_t$, add a new CNOT gate after), defined as
\begin{equation}
    (g_1, \ldots, g_t, g_{t+1}, \ldots, g_n) \mapsto (g_1, \ldots, g_t, g_\text{new}, g_{t+1}, \ldots, g_n).
    \label{eq:gate_addition}
\end{equation}
The second action is gate removal (${ \bf p_2 = p_\text{remove} }$), where the action omits $g_t$ from a sequence of gates (e.g., removing a $X$-rotation gate), defined as 
\begin{equation}
    (g_1, \ldots, g_{t-1}, g_t, g_{t+1}, \ldots, g_n) \mapsto (g_1, \ldots, g_{t-1}, g_{t+1}, \ldots, g_n).
    \label{eq:gate_removal}
\end{equation}
Switching the gate type is the third possible action (${ \bf p_3 = p_\text{switch}}$), where it replaces $g_t$ with a different gate type acting on the same number of wires (e.g., $X$-rotation replaced with $Y$-rotation or CNOT replaced with CRX gate), defined as
\begin{equation}
    (g_1, \ldots, g_t, \ldots, g_n) \mapsto (g_1, \ldots, g_t', \ldots, g_n)
    \label{eq:gate_switch}
\end{equation}
The last gate action is a gate move (${ \bf p_4 = p_\text{move}}$), where the action moves $g_t$ to act on different qubits (e.g., move $X$-rotation from qubit $1$ to qubit $3$), defined as
\begin{equation}
    (g_1, \ldots, g_t, \ldots, g_n) \mapsto (g_1, \ldots, g_t', \ldots, g_n)
    \label{eq:gate_move}.
\end{equation}

\noindent During an iteration of the sampling process, the choice of $p_i$ enables LQAS to limit the expected number of actions performed on an ansatz.
This permits users to bound the local search space and the speed of exploration of the space by restricting how much, on average, each sampled circuit deviates from its base circuit on each iteration.
For example, if an ansatz $A$ has $n$ modifiable gates, then an average of $k$ gate addition actions, where $k < n$, can be imposed by setting $p_\text{add} = \frac{k}{n}$.

The sampling process of the LQAS algorithm is controlled by two parameters that dictate the size of the baseline population and the number of modified circuits to collect from each. 
The number of best-performing circuits chosen for further resampling of candidate circuits in the next iteration, is denoted as $\bm{K}$, and the number of new circuits sampled from each of the top circuits in the iteration is defined as $\bm{N}$.
To perform modification sampling, LQAS iterates through an ansatz $A$ and samples a Bernoulli random variable $X_i$ for each $i \in \{P\}$ for each gate $g_t \in A$, where $P$ is the set of gate modifications.
The Bernoulli distribution is parametrized by the action probabilities.
If $X_i = 1$ for the action $i$ on the gate $g_t$, then the action $i$ is applied to the gate, and no further modifications are sampled for this iteration. 
If $X_i = 0$, then action $i$ is not applied to the gate.
Any particular gate modification action represented by a subscript $i$ is defined as $X_i \sim \text{Bernoulli}(p_i)$.

\begin{figure}[H]
    \centering

    \includegraphics[width=\linewidth]{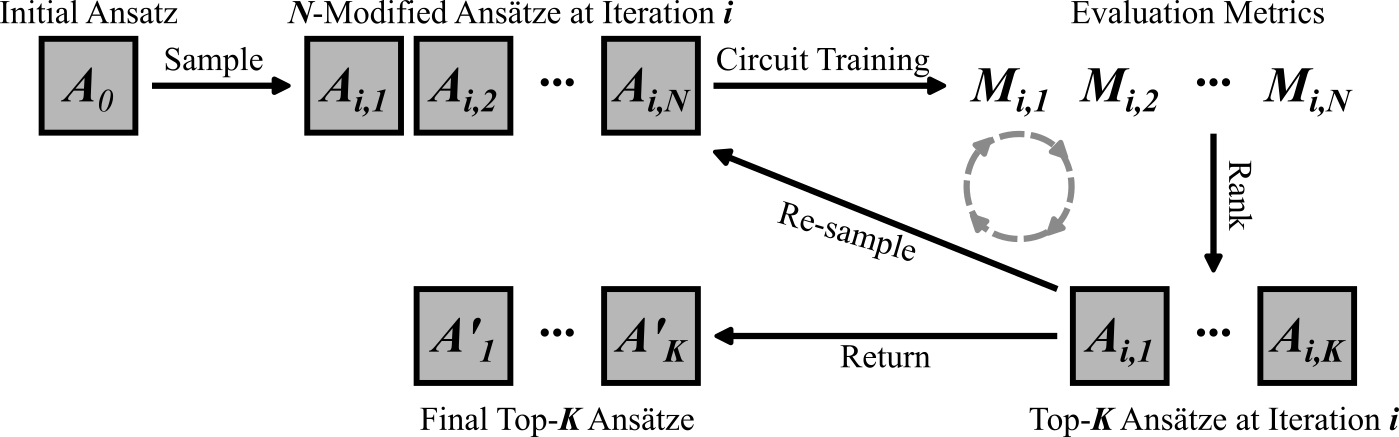}
    \caption{The flow diagram above demonstrates the iterative ansatz modification process in the LQAS framework.}
    \label{fig:combined-sampling}
\end{figure}

In Figure~\ref{fig:combined-sampling}, we provide a flow diagram demonstrating the ansatz modification process.
Starting with an initial ansatz, $A_{0}$, the sampling process generates a set of $\bm{N}$-ans\"{a}tze, defined as $A_{i,1}, \ldots, A_{i,\bm{N}}$, where $i$ denotes the $i$th-iteration of the LQAS process.
Therefore, the initial set of $\bm{N}$-modified ans\"{a}tze is defined as $A_{0,1}, \ldots, A_{0,\bm{N}}$. 
This set of circuits is then evaluated on the given learning task and ranked using a user-defined metric, such as the validation loss.
For each circuit, we define the set of evaluation metrics as $M_{i,1}, \ldots, M_{i,\bm{N}}$, where $M_{i,1}$ would be the evaluation metric for circuit $A_{i,1}$.
This set of evaluation metrics is then used to rank the candidate ans\"{a}tze, where we choose the $\bm{N}$ circuits to be re-sampled and passed through the modification process using a user-defined set of iterations.
Following the termination of the modification process, the final set of circuits is defined as $A_{1}^{\prime}, \ldots, A_{\bm{K}}^{\prime}$.
Lastly, the performance between the initial ansatz, $A_{0}$, and final ``best'' circuit, $A_{1}^{\prime}$, can be compared to highlight the performance increase between the initial and final ans\"{a}tze.

\subsection{Datasets}
\label{subsection:datasets}
To study the effects of local gate modification in the LQAS algorithm, we explore several synthetic function-fitting datasets through empirical evaluations.
Regardless of how well the baseline circuit can model a function, we evaluate whether LQAS produces meaningful changes to the ansatz that enable it to model arbitrary linear and non-linear functions.
In these datasets, we include one- and two-dimensional quadratic functions embedded with noise drawn independently and identically distributed (i.i.d.) from a normal distribution to understand the searching algorithms' capacity to find circuits capable of handling inherently non-uniform and noisy data.
The first synthetic dataset is generated by a quadratic function $f:(-2, 2) \rightarrow \mathbb{R}$. 
\begin{equation}
    \label{eq:1d-noisy}
    f(x) = x^2 + \epsilon
\end{equation}
\noindent where $x \in (-2, 2)$ and $\epsilon \sim \mathcal{N}(0, 0.5)$. 500 datapoints were generated with i.i.d samples of $\epsilon$ split into 80\% training set and 20\% validation. Similarly, the second synthetic dataset is generated by the function $g:\left( -1, 1 \right) \times \left( -1, 1 \right) \rightarrow \mathbb{R}$.
\begin{equation}
    \label{eq:2d-noisy}
        g(x,y) = x^2 + y^2 + \epsilon
\end{equation}
\noindent where $\epsilon \sim \mathcal{N}(0, 0.5)$.
We restrict $x,y \in \left( -1, 1 \right)$ and generate 200 data points using these parameters which were split into 80\% training set and 20\% validation. 

The first quantum chemistry dataset used for LQAS evaluation is generated in the framework of the data-driven coupled-cluster (DDCC) scheme~\cite{Townsend2019-gx,Jones2023-hz}.
DDCC is a classical machine learning-based approach for predicting $t_{2}$-amplitudes, i.e., two-electron excitation amplitudes, of the coupled-cluster singles and doubles (CCSD) wave function.
The goal of this method is to reduce the number of iterations to converge the coupled-cluster equations using features generated from lower-level methods used to initialize the CCSD wave function, such as Hartree-Fock (HF) and M{\o}ller-Plesset second-order perturbation theory (MP2).
The dataset used throughout this study consists of 199 water conformers obtained from the study of Townsend and Vogiatzis~\cite{Townsend2019-gx}.
All electronic structure calculations (i.e., HF, MP2, and CCSD calculations) were performed using Psi4~\cite{parrish_psi4_2017} and Psi4Numpy~\cite{smith_psi4numpy_2018} with the STO-3G basis set~\cite{hehre_selfconsistent_1970} and the frozen core approximation.
Despite this dataset consisting of simple molecules, generated using a minimal basis set, this is a data-intensive quantum chemistry application for PQCs since each molecule has $(N_{occ})^{2}(N_{virt})^{2}$ $t_{2}$-amplitudes, where $N_{occ}$ denotes the number of occupied orbitals and $N_{virt}$ denotes the number of virtual (or unoccupied) orbitals.
In our dataset, there are 4 occupied and 2 virtual orbitals for the STO-3G basis set of each water conformer, which corresponds to a total of 64 $t_{2}$-amplitudes per conformer.

Since the PQCs we examine in this study employ a direct one-to-one mapping between features and qubits, the initial DDCC feature set of 30 features can be compressed using dimensionality-reduction techniques.
Another motivation for dimensionality reduction is that wider quantum circuits result in greater depth, which makes them more costly to evaluate using state-vector simulation and real hardware.
The feature reduction for this feature set is analyzed in~\citet{Jones2025-kt}, where the original set of 30 features is reduced to five features (for a five-qubit system) using SHapley Additive exPlanation (SHAP)~\cite{Kumar2020-ub}. SHAP uses an explanatory additive model to analyze features in a dataset based on concepts from cooperative game theory. 
From the SHAP analysis,  the top five features correspond to the two-electron integrals ($\mel{ij}{}{ab}$), MP2 $t_{2}$-amplitudes ($t^{ab}_{ij(\text{MP2})}$), the magnitude of the MP2 $t_{2}$-amplitudes, the difference in orbital energies ($\varepsilon_{i}+\varepsilon_{j}-\varepsilon_{a}-\varepsilon_{b}$), and the binary feature denoting whether two-electrons are promoted to the same virtual orbital ($a=b$).
Additional details regarding the DDCC are provided in Appendix Section~\ref{section:dataset_ddcc}.

The second quantum chemistry dataset we explore in this study is the BSE49 dataset~\cite{Prasad2021-sx}, which consists of molecular structures and bond separation energies that are useful for benchmarking ML methods in computational chemistry.
The BSE49 dataset consists of 49 unique A-B covalent bonds, which represent homolytic bond cleavages of 4394 molecular structures; all of which are calculated using the restricted-open-shell complete basis set-quadratic Becke3 ((RO)CBS-QB3)~\cite{Wood2006-zn, Montgomery1999-fr, Montgomery2000-cn} composite quantum chemistry method.
Of the 4394 structures in the dataset, 1951 of them correspond to existing structures, while the remaining 2443 are hypothetical structures.
Like the DDCC dataset, we use this dataset in our previous work (\citet{Jones2025-kt}), where we analyze various molecular representations and feature reduction techniques to find the optimal combination.
During this process, we found that seven of the hypothetical structures needed to be removed during the preprocessing stage due to valency (bonding) exception issues when converting Cartesian coordinates to \textsc{RDKit}~\cite{Landrum2023-ov} \textsc{mol} objects.
We found that Morgan (or extend-connectivity) fingerprints~\cite{Morgan1965-kv, Rogers2010-yp}, which are based on encoding structural details from traversals of the molecular graph into bit vectors, performed best when paired with principal component analysis~\cite{Mackiewicz1993-fc}, a linear dimensionality reduction technique, to reduce the feature set from 2048 features to 16 features.
Additionally, we found that representing the feature vector as the deviation between the product and reaction feature vectors, i.e., $\mathbf{X}_{\mathit{sub}} = (\mathbf{X}_{\text{A}^{.}} + \mathbf{X}_{\text{B}^{.}}) - \mathbf{X}_{\text{A-B}}$ (similar to the method used in the study of~\citet{garcia-andrade_barrier_2023}), offers favorable performance over the reactants (i.e., $\mathbf{X}_{\text{A-B}}$) alone.
For more information regarding the dataset preprocessing, the reader is referred to our previous work ~\cite{Jones2025-kt}.

Lastly, we will provide a few brief comments on our choice of chemically relevant datasets.
While access to quantum chemistry datasets has become more readily available to the machine learning in chemistry community through resources, like the Molecular Software Science Institute (MolSSI) Machine Learning Datasets Repository~\cite{molssiMolSSIMLDatasets} and datasets available on \href{https://quantum-machine.org/}{https://quantum-machine.org/}~\cite{quantummachineQuantumMachineorgDatasets}, we wanted to explore domain-specific problems that offer unique insights into the performance of PQCs.
The DDCC dataset offers a learning task that is simple for classical ML algorithms, while being challenging for QML algorithms due to large number of samples in the dataset.
In this work, we show (\textit{vide infra}) that this dataset can be used for benchmarking due to challenges related to real device execution.
Alternatively, the BSE49 dataset can also provide unique insights into challenges that arise when mapping traditional molecular representations to quantum circuits.
This dataset can also provide unique insights into the challenges of using a compressed set of features with state-vector simulations.
Overall, we argue that, while neither dataset has been broadly applied to benchmark QML approaches, they can serve as useful, challenging benchmarking edge cases for the community.

\subsection{Computational Details}
\label{subsection:compdetails}
To aid in the development, evaluation, and experimentation of the LQAS method, we developed the algorithm using two common quantum computing software development kits (SDKs), \textsc{Pennylane}~\cite{Bergholm2018-ex} and \textsc{Qiskit}~\cite{Javadi-Abhari2024-nw}.
These two SDKs are used in tandem since we first implemented the LQAS algorithm in \textsc{Pennylane} and then used the \textsc{Qiskit} code from our earlier work~\cite{Jones2025-kt} to run an initial and final circuit from a \textsc{Pennylane}-based LQAS experiment on real quantum hardware, including \textit{ibm{\textunderscore}fez} and \textit{ibm{\textunderscore}kingston}, both of which are 156-qubit Heron r2 quantum processing units.
As mentioned in ~\citet{Jones2025-kt}, the primary motivation for implementing our code in \textsc{Qiskit} is to use the \textsc{Qiskit} Batch Execution mode, which is not available in the \textsc{PennyLane-Qiskit} plugin. 

For all models, features ($\mathbf{X}$) and target values ($\mathbf{y}$) were scaled on $\mathbb{R}\in [ -1,1 ]$ by applying the \textsc{MinMaxScaler}, as implemented in \textsc{Scikit-learn}~\cite{pedregosa_scikit-learn_2011}. 
All state-vector simulations were performed with \textsc{Pennylane} using the \textsc{Qulacs}~\cite{Suzuki2020-pc} backend, with circuit parameters initialized to 0, and optimized using the Adam optimizer~\cite{Kingma2014-tz} for 200 iterations with a batch size of 25 and a learning rate of $10^{-2}$. 
The circuits ran in \textsc{Qiskit} utilized 500 training iterations, the constrained optimization by linear approximation (COBYLA) optimizer implemented in \textsc{SciPy}\cite{virtanen_scipy_2020}, and the circuit parameters were initialized on $\mathbb{R} \in [-\pi, \pi]$.

For experiments on real quantum hardware, we adapted the \textsc{Qiskit} code from~\citet{Jones2025-kt}, which utilizes the \textsc{Qiskit} Batch Execution mode.
This is important since on real hardware, data-intensive models, such as the DDCC models that contain many samples (i.e., $N_{molecules} \times N_{occ}^{2} \times N_{virt}^{2} = 199 \times 4^2 \times 2^2 = 12,736$ total samples) are not able to run using the \textsc{Qiskit} Session Execution mode provided by the \textsc{PennyLane-Qiskit} plugin.
Additionally, experiments using noisy simulation were performed using \textit{qiskit-aer} with the \textit{FakeQuebec} backend to benchmark circuit optimization and resilience levels.
We found that the maximum optimization level, related to \textsc{Qiskit}’s transpiler settings, corresponded to level one, a light optimization level that incorporates layout optimization, inverse cancellation of sequential gates that are inverses of each other, and one-qubit gate optimization.
The maximum resilience, or error mitigation, was set to level one, which corresponds to Twirled Readout Error eXtinction (TREX)~\cite{van_den_berg_model-free_2022} error mitigation.
Lastly, the maximum number of shots, or device executions, using these optimization and resilience levels was 3072.

To evaluate the effectiveness of the selected ansatz and produce QML circuits by quantifying model fitness, we chose two metrics implemented in \textsc{scikit-learn}~\cite{Pedregosa2011-bf}.
The first is the mean-squared error (MSE) and the second is the coefficient of determination ($R^2$).
The motivation for choosing two evaluation metrics is as follows: the MSE measures the average deviation of predicted values from their corresponding ground truth, while the $R^{2}$ quantifies how much of the variation in the target values is explained by the model's predictions.
Therefore, an ideal model has an MSE near $0$ and an $R^2$ close to 1.

\section{Results}\label{section:Results}
As mentioned in Section \ref{subsection:datasets}, to explore how LQAS designs circuits for QML regression-based tasks, we evaluate our method on synthetic and chemical datasets.
Starting with the synthetic function fitting datasets, we observe that LQAS yields circuits with significantly better regression performance than the baseline ansatz. 
Following these observations, we note that for both chemical datasets (DDCC and BSE49), model performance improves throughout the LQAS iterations and is consistent across training and validation datasets.
Additionally, based on these results, we note that the model produced by the final iteration of LQAS is not necessarily better than models produced earlier, and that the difference in performance is greatest between the baseline circuit and the first iteration circuit of LQAS.

\subsection{Synthetic Datasets}

The first synthetic dataset we analyze is the one-dimensional quadratic function generated by Equation \ref{eq:1d-noisy}. 
We examine the performance of LQAS on this dataset using nine different base ans\"{a}tze including HEA-k-n where $k\ge n$ for $k = 1, 2, 3$. 
The performance of all nine base circuits is highlighted in Table \ref{tab:noisy-1Dquadratic}.
Due to the dimensionality of the input vector for each sample, we explore a four-qubit system where a single input value is encoded onto each qubit. 
We run the LQAS algorithm for three iterations, sampling 100 circuits each iteration, and select the top ten to produce the base population for the next iteration. 
For gate modifications, we evaluate LQAS with $p_\text{add} = p_\text{remove} = p_\text{switch} = p_\text{move} = 0.1$, giving each type of modification a 10\% chance of occurring for each gate in the circuit. 

\begin{figure*}[t]
\centering
\begin{minipage}[t]{0.49\textwidth}
\centering
\begin{tikzpicture}[scale=1]
\begin{axis}[
    xlabel={Iteration},
    ylabel={MSE (unitless)},
    xtick={0,1,2,3},
    xticklabels={Base, 1, 2, 3},
    legend pos=north east,
    ymajorgrids,
    grid style=dashed,
    ymin=-0.5, ymax=3,
    width=0.95\linewidth,
    height=0.55\linewidth,
    title={MSE Across Iterations},
]
\addplot[
    color=red,
    mark=*,
    thick,
    nodes near coords,
    point meta=explicit symbolic,
] coordinates {
    (0, 2.5339) [2.534]
    (1, 0.0618) [0.062]
    (2, 0.0698) [0.070]
    (3, 0.0537) [0.054]
};
\addlegendentry{MSE (↓ better)}
\end{axis}
\end{tikzpicture}
\end{minipage}
\hfill
\begin{minipage}[t]{0.49\textwidth}
\centering
\begin{tikzpicture}[scale=1]
\begin{axis}[
    xlabel={Iteration},
    ylabel={$R^2$},
    xtick={0,1,2,3},
    xticklabels={Base, 1, 2, 3},
    legend pos=south east,
    ymajorgrids,
    grid style=dashed,
    ymin=-1.1, ymax=1.3,
    width=0.95\linewidth,
    height=0.55\linewidth,
    title={$R^2$ Across Iterations},
]
\addplot[
    color=blue,
    mark=square*,
    thick,
    nodes near coords,
    point meta=explicit symbolic,
] coordinates {
    (0, -0.9928) [-0.993]
    (1, 0.9514) [0.951]
    (2, 0.9451) [0.945]
    (3, 0.9577) [0.958]
};
\addlegendentry{$R^2$ (↑ better)}
\end{axis}
\end{tikzpicture}
\end{minipage}
\caption{Model evaluation of HEA-1-1 using the one-dimensional noisy quadratic function performance for the best model from each iteration over 3 iterations. Overall, iteration 3 yields the best function approximation, lowest MSE, and highest $R^2$.}
\label{fig:noisy-quadratic-performance}
\end{figure*}
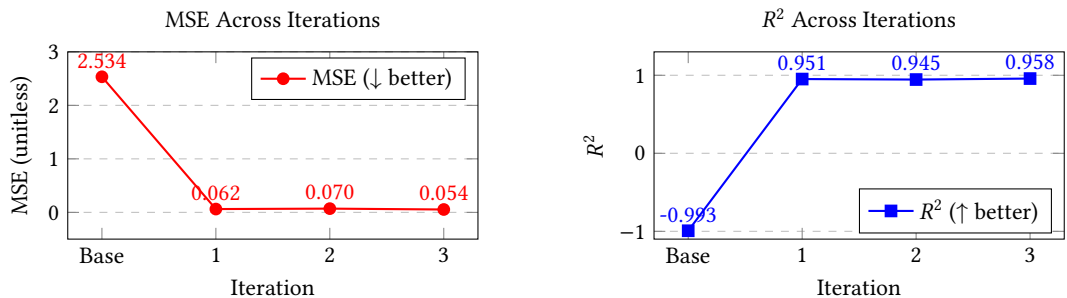

Starting with HEA-1-1, the simplest model, we found that after training, the base model was unable to fit the function properly, resulting in a validation $R^2$ of $-0.993$ and an MSE of 2.534. 
Despite the initial negative $R^2$, the model performance can be further improved using the LQAS algorithm.
The modified circuit provided by LQAS after three iterations has a final $R^2$ of 0.958 and an MSE of 0.054, as demonstrated in Figure~\ref{fig:noisy-quadratic-performance}.
In this setting, we found that the greatest improvement occurred after the first iteration, which then gradually improved with the second and third iterations to the final value.
This emphasizes the exploratory nature of LQAS, making it a valuable tool for systematically traversing the architecture decision space.

\begin{figure*}[ht!]
    \centering
    \begin{minipage}[b]{0.48\textwidth}
        \centering
        \resizebox{\textwidth}{!}{
        \begin{quantikz}
            \lstick{0} & \gate{R_X} & \gate{R_Y} & \gate{R_Z} & \gate{R_Y} & \ctrl{1} & \qw      & \qw      & \qw & \targ{0}  & \gate{R_X} & \gate{R_Y} & \gate{R_Z} & \gate{R_Y} & \ctrl{1} & \qw      & \qw      & \qw & \targ{0} & \meter{} \\
            \lstick{1} & \gate{R_X} & \gate{R_Y} & \gate{R_Z} & \gate{R_Y} & \targ{}  & \ctrl{1} & \qw      & \qw & \qw      & \gate{R_X} & \gate{R_Y} & \gate{R_Z} & \gate{R_Y} & \targ{}  & \ctrl{1} & \qw      & \qw & \qw & \qw \\
            \lstick{2} & \gate{R_X} & \gate{R_Y} & \gate{R_Z} & \gate{R_Y} & \qw      & \targ{}  & \ctrl{1} & \qw & \qw      & \gate{R_X} & \gate{R_Y} & \gate{R_Z} & \gate{R_Y} & \qw      & \targ{}  & \ctrl{1} & \qw & \qw & \qw \\
            \lstick{3} & \gate{R_X} & \gate{R_Y} & \gate{R_Z} & \gate{R_Y} & \qw      & \qw      & \targ{}  & \qw & \ctrl{-3} & \gate{R_X} & \gate{R_Y} & \gate{R_Z} & \gate{R_Y} & \qw      & \qw      & \targ{}  & \qw & \ctrl{-3} & \qw
        \end{quantikz}
        }
    \end{minipage}
    \hfill
    \begin{minipage}[b]{0.48\textwidth}
        \centering
        \resizebox{\textwidth}{!}{
        \begin{quantikz}
            \lstick{0} & \gate{R_X} & \gate{R_Y} & \gate{R_Z} & \gate{R_Y} & \qw     & \qw    & \gate{R_Z} & \gate{R_X} & \gate{R_Y} & \gate{R_Z} & \gate{R_Y} & \ctrl{1} & \qw      & \qw      & \qw & \targ{0} & \meter{} \\
            \lstick{1} & \gate{R_X} & \gate{R_Y} & \gate{R_Z} & \gate{R_Y} & \ctrl{1} & \qw & \qw      & \gate{R_X} & \gate{R_Y} & \gate{R_Z} & \gate{R_Y} & \targ{}  & \ctrl{1} & \qw      & \qw & \qw & \qw \\
            \lstick{2} & \gate{R_X} & \gate{R_Y} & \gate{R_Z} & \gate{R_Y} & \targ{} & \ctrl{1} & \qw & \gate{R_X} & \gate{R_Y} & \gate{R_Z} & \gate{R_Y} & \qw      & \targ{}  & \ctrl{1} & \qw & \qw & \qw \\
            \lstick{3} & \gate{R_X} & \gate{R_Y} & \gate{R_Z} & \gate{R_Y} & \qw     & \targ{}    & \ctrl{-3} & \gate{R_X} & \gate{R_Y} & \gate{R_Z} & \gate{R_Y} & \qw      & \qw      & \targ{}  & \qw & \ctrl{-3} & \qw
        \end{quantikz}
        }
    \end{minipage}
    \caption{Comparison of the base circuit (left) and optimized circuit after three iterations of LQAS (right).}
    \label{fig:1d-quad:circuit-comparison}
\end{figure*}
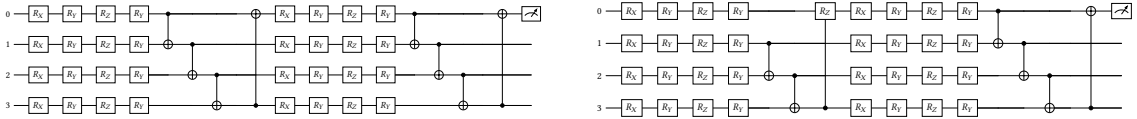

Similarly, Figure~\ref{fig:1d-quad:circuit-comparison} highlights how minimal modifications of the base HEA-1-2 model can improve model performance.
While the initial model has an $R^2$ of 0.902 and an MSE of 0.125, the final model, which includes the removal of a CNOT gate, the replacement of a CNOT gate with a CRZ gate, shows an increase in $R^2$ to $0.993$ and a $94\%$ reduction in MSE from $0.125$ to $0.008$. 
This is achieved by making minimal modifications to the baseline ansatz, as shown in Figure~\ref{fig:1d-quad:circuit-comparison}, which further demonstrates the use of pre-selected ans\"{a}tze as baseline templates and the potential impact of small local modifications on regression accuracy.

For this dataset, we highlight the overall performance of all nine models in Table \ref{tab:noisy-1Dquadratic}.
The base models have a minimum $R^2$ of -0.993, a maximum $R^2$ of 0.992, and a mean $R^2$ of 0.746.
Across the three subsequent LQAS iterations, the minimum $R^2$ improves to 0.94 or above; specifically, the minimum $R^2$ values correspond to  0.951, 0.945, and 0.958, respectively.
The maximum $R^2$ values indicate improved performance across all three iterations compared to the base model, with maximum values of 0.998, 0.998, and 0.997, respectively.
Like the minimum and maximum $R^2$ values, the mean $R^2$ values also show improvement across the LQAS iterations.
The mean value improves from an $R^2$ of 0.746 to 0.988, 0.988, and 0.990 in the first, second, and third LQAS iterations, respectively.
We note that the minimum values from the base model and the three LQAS models correspond to the previously highlighted HEA-1-1 model from Fig. \ref{fig:noisy-quadratic-performance}, while the maximum values correspond to HEA-2-3, HEA-3-2, HEA-3-2, HEA-3-3, for the base, first, second, and third LQAS iteration, respectively.
Both the minimum and maximum values highlight the robustness of the LQAS model, where the minimally expressive model (or HEA-1-1) demonstrates significant improvements, while the more expressive models (HEA-2-3, HEA-3-2, HEA-3-3) demonstrate marginal improvements throughout the circuit optimization process.

The second synthetic dataset we use is a two-dimensional quadratic function generated by Equation \ref{eq:2d-noisy}.
Similar to the one-dimensional quadratic function (Equation \ref{eq:1d-noisy}), we analyze multiple base ans\"{a}tze including HEA-1-1, HEA-1-2, HEA-1-3, HEA-1-4, HEA-2-2, HEA-3-2, HEA-4-2.
Due to the performance of the one-dimensional noisy quadratic, which implemented redundant encodings, for the two-dimensional noisy quadratic, the $x$ and $y$ inputs each are mapped to a single qubit to form a two-qubit circuit.
For this dataset, we run the algorithm for three iterations, sample 100 circuits each iteration, and select the top ten to produce the baseline population for the next iteration.
For gate modifications, we evaluate LQAS with $p_\text{add} = p_\text{remove} = p_\text{switch} = p_\text{move} = 0.1$, giving each type of modification a 10\% chance of occurring for each gate in the circuit.
By design, this does not form a single complete distribution, so not every gate in the circuit is modified.
Instead, for an arbitrary two-dimensional function, we choose these probabilities to ensure an equal chance of occurrence.

The results are highlighted in Table \ref{tab:noisy-2Dquadratic}, where the base models for the seven circuits explored on this dataset have a minimum $R^{2}$ value of -0.520 (HEA-1-1), a maximum $R^{2}$ value of 0.888 (HEA-1-3), and an average $R^{2}$ value of 0.415.
Like the previous dataset, these models show improvements throughout the LQAS process.
Starting with the first iteration, which has a minimum value $R^{2}$ of -0.207 (HEA-1-1), maximum value $R^{2}$ of 0.960 (HEA-1-4), and mean value of 0.771.
This shows an improvement of 0.313 for the minimum value, 0.072 for the maximum value, and 0.356 for the mean value.
The performance of the LQAS algorithm on this dataset is further highlighted by the deviations from the last LQAS iteration and the base model.
The third (final) iteration of LQAS has a minimum value $R^{2}$ of 0.137 (HEA-1-1), a maximum value $R^{2}$ of 0.958 (HEA-3-2), and a mean value of 0.835.
Due to using a one-to-one input to qubit encoding, this model demonstrates drastic improvements between the base and final LQAS iteration.
Even for the worst performing ansatz, HEA-1-1, the model shows a drastic improvement regarding the $R^{2}$, which increases by 0.657 from the base model to final LQAS iteration.
Similarly, the remaining models all improve to $R^{2}$s greater than 0.93 for the final iteration, further highlighting the improvements offered by the LQAS algorithm. 

\subsection{Quantum Chemistry Datasets}
Following the demonstration of LQAS's ability to model arbitrary noisy functions, we examine the method's performance on two quantum chemistry datasets.
Starting with the DDCC dataset, we demonstrate LQAS's aptitude to model chemical regression tasks.
We run the algorithm for five iterations, sampling 100 circuits each iteration, and select the top ten to produce the base population for the next iteration.
For gate modifications, we evaluate LQAS with $p_\text{add} = p_\text{remove} = p_\text{switch} = p_\text{move} = 0.1$, giving each type of modification a 10\% chance of occurring for each gate in the circuit.
We determined that equal probabilities for each sampling action were adequate based on the performance of the previously modeled noisy quadratic functions.

Due to the cost of state-vector simulation on a regression-task that contains a large number of samples, we restrict our model search to a smaller subset of HEA circuits.
Additionally, due to this restriction, preliminary analysis showed that the HEA-3-2 circuit offered a balanced model regarding accuracy and computational cost.
Therefore, based on these considerations, we only analyze the HEA-3-2 circuit for this dataset.
This is highlighted in Table~\ref{tab:ddccmodel} in the Appendix, where we present the loss values for the training and validation data of the best models from each iteration of the LQAS algorithm.
Note that the loss values are unitless due to the regression target value being a coefficient (or amplitude) that weights the two-electron excitation.
The base model has an $R^{2}$ of 0.946 for the training set and 0.948 on the validation set, which is further improved over the subsequent LQAS iterations with $R^{2}$s ranging from 0.986-0.993 for the training set and 0.987-0.993 for the validation set.
This demonstrates LQAS's ability to accurately design a PQC to model the DDCC dataset and generalize well to unseen data due to the lack of overfitting between the training and validation set.

\begin{figure*}[t]
\centering
\begin{tikzpicture}
\begin{axis}[
    xlabel={Iteration},
    ylabel={Loss (log scale; unitless)},
    xtick={0,1,2,3,4,5},
    xticklabels={Base, 1, 2, 3, 4, 5},
    legend style={at={(0.98,0.5)}, anchor=east},
    ymajorgrids=true,
    grid style=dashed,
    ymode=log,
    ymin=1e-4, ymax=1,
    width=0.95\linewidth,
    height=0.35\linewidth,
    title={Train and validation loss},
    every node near coord/.append style={font=\scriptsize},
    nodes near coords align={vertical},
]

\addplot[
    color=plotgreen,
    mark=*,
    thick,
    nodes near coords,
    point meta=explicit symbolic,
] coordinates {
    (0, 0.3557) [0.356]
    (1, 0.2355) [0.236]
    (2, 0.0618) [0.062]
    (3, 0.1773) [0.177]
    (4, 0.1082) [0.108]
    (5, 0.1334) [0.133]
};
\addlegendentry{Train}

\addplot[
    color=plotred,
    mark=square*,
    thick,
    nodes near coords,
    point meta=explicit symbolic,
] coordinates {
    (0, 0.0047) [0.0047]
    (1, 0.0012) [0.0012]
    (2, 0.0008) [0.0008]
    (3, 0.00095) [0.00095]
    (4, 0.0008) [0.0008]
    (5, 0.0006) [0.0006]
};
\addlegendentry{Validation}

\end{axis}
\end{tikzpicture}
\caption{
DDCC regression performance over LQAS iterations.
Panel (a) shows log-scaled train/validation loss; panel (b) lists the corresponding metrics.
}
\label{fig:ddcc-tabular}
\end{figure*}
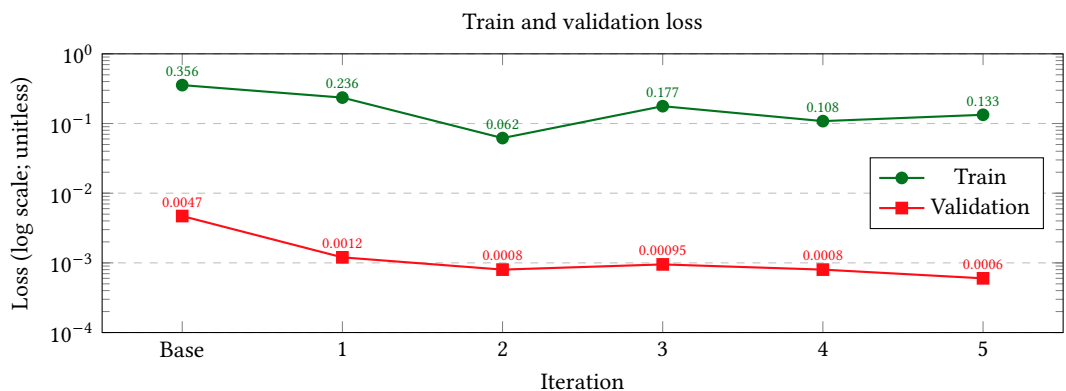

Following the evaluation of the DDCC dataset, BSE49 is the last dataset we examine in this study.
BSE49 represents a common approach for machine learning in chemistry regression tasks where the Cartesian coordinates of a molecule are mapped to a molecular representation and the regression target is a molecular property, in this case, bond separation energies.
As mentioned in Section \ref{subsection:datasets}, for this dataset, we use a 16-dimensional feature vector that corresponds to a 16-qubit mapping.
In this setting, we deploy LQAS on a large 16-qubit HEA-3-3 ansatz that has a total of 192 trainable parameters.
For this circuit, gate addition is favored, whereas gate movement is discouraged, to preserve the overall HEA structure as much as possible.
As such, the sampling probabilities are set to $p_\text{add} = 1/12$, $p_\text{remove} =  p_\text{switch} = 1/24$, and $p_\text{move} = 1/48$.
On this dataset, we run the LQAS algorithm for ten iterations, using 100 samples in each iteration, while selecting the top 30 to produce the baseline population for the next iteration. 

As presented in Figure~\ref{fig:combined_bse16}, the base model has an $R^{2}$ of 0.314 and an MSE of 0.148 for the training set and an $R^{2}$ of 0.348 and an MSE of 0.072 for the validation set.
Despite the $R^{2}$ of the training and validation sets improving to values of 0.425 and 0.456, respectively, throughout the LQAS process on this dataset, the performance improvement between the base model and subsequent LQAS iterations is marginal.
The plateau in performance, with $R^{2}$ values ranging from 0.425-0.502, could be explained by model capacity and data compression.
As mentioned in~\citet{Jones2025-kt}, the BSE49 dataset is very non-linear and contains a wide range of BSE values.
Additionally, this task requires reducing a 2048-dimensional feature vector down to 16 features.
This small set of features, along with the task of learning a non-linear dataset using only 192 trainable parameters, could result in a model that is not flexible enough for this complex dataset.
Additionally, due to the cost of state-vector simulation on this HEA-3-3 model, further evaluations using more complex circuits were not performed.

\begin{figure*}[t]
\centering

    \begin{minipage}[b]{0.45\textwidth}
        \centering

        \begin{tikzpicture}
        \begin{axis}[
            xlabel={Iteration},
            ylabel={Loss ((kcal/mol)$^2$)},
            xtick={0,1,...,9},
            xticklabels={Base, 1, 2, 3, 4, 5, 6, 7, 8, 9},
            legend style={at={(1.0,0.5)}, anchor=east},
            ymajorgrids=true,
            grid style=dashed,
            ymin=0, ymax=0.23,
            width=\linewidth,
            height=0.7\linewidth,
            title={Training and Validation Loss across Iterations},
            every node near coord/.append style={font=\scriptsize},
            nodes near coords align={vertical},
        ]
        
        \addplot[
            color=plotred,
            mark=*,
            thick,
            nodes near coords,
            point meta=explicit symbolic,
        ] coordinates {
            (0, 0.1477) [0.148]
            (1, 0.1178) [0.118]
            (2, 0.1889) [0.189]
            (3, 0.1690) [0.169]
            (4, 0.2031) [0.203]
            (5, 0.1498) [0.150]
            (6, 0.1667) [0.167]
            (7, 0.2141) [0.214]
            (8, 0.1883) [0.188]
            (9, 0.0332) [0.033]
        };
        \addlegendentry{Train Loss}
        
        \addplot[
            color=plotgreen,
            mark=square*,
            thick,
            nodes near coords,
            point meta=explicit symbolic,
            nodes near coords style={anchor=south},
        ] coordinates {
            (0, 0.0721) [0.072]
            (1, 0.0602) [0.060]
            (2, 0.0571) [0.057]
            (3, 0.0568) [0.057]
            (4, 0.0564) [0.056]
            (5, 0.0550) [0.055]
            (6, 0.0556) [0.056]
            (7, 0.0565) [0.057]
            (8, 0.0570) [0.057]
            (9, 0.0551) [0.055]
        };
        \addlegendentry{Validation Loss}
        
        \end{axis}
        \end{tikzpicture}
        \caption*{\textbf{(a)}}
        \label{fig:loss_bse16}
    \end{minipage}
    \hfill
    \begin{minipage}[b]{0.45\textwidth}
        \centering
        \begin{tikzpicture}
        \begin{axis}[
            xlabel={Iteration},
            ylabel={$R^2$},
            xtick={0,1,...,9},
            xticklabels={Base, 1, 2, 3, 4, 5, 6, 7, 8, 9},
            legend style={at={(1.0,0.05)}, anchor=south east},
            ymajorgrids=true,
            grid style=dashed,
            ymin=0.30, ymax=0.55,
            width=\linewidth,
            height=0.7\linewidth,
            title={Training and Validation $R^2$ across Iterations},
            every node near coord/.append style={font=\scriptsize},
            nodes near coords align={vertical},
        ]
        
        \addplot[
            color=plotred,
            mark=*,
            thick,
            nodes near coords,
            point meta=explicit symbolic,
        ] coordinates {
            (0, 0.3134) [0.31]
            (1, 0.4251) [0.43]
            (2, 0.4254) [0.43]
            (3, 0.4411) [0.44]
            (4, 0.4409) [0.44]
            (5, 0.4387) [0.44]
            (6, 0.4431) [0.44]
            (7, 0.4064) [0.41]
            (8, 0.4388) [0.44]
            (9, 0.4296) [0.43]
        };
        \addlegendentry{Train $R^2$}
        
        \addplot[
            color=plotgreen,
            mark=square*,
            thick,
            nodes near coords,
            point meta=explicit symbolic,
            nodes near coords style={anchor=south},
        ] coordinates {
            (0, 0.3483) [0.35]
            (1, 0.4555) [0.46]
            (2, 0.4840) [0.48]
            (3, 0.4868) [0.49]
            (4, 0.4902) [0.49]
            (5, 0.5029) [0.50]
            (6, 0.4978) [0.50]
            (7, 0.4889) [0.49]
            (8, 0.4848) [0.48]
            (9, 0.5018) [0.50]
        };
        \addlegendentry{Validation $R^2$}
        
        \end{axis}
        \end{tikzpicture}
        \caption*{\textbf{(b)}}
        \label{fig:r2_bse16}
    \end{minipage}

    \caption{Performance metrics of the best model across LQAS iterations on the BSE49 dataset. minipages show (a) MSE trends, (b) $R^2$ trends, and (c) training/validation statistics.}
    \label{fig:combined_bse16}
\end{figure*}

\subsection{LQAS on Real Quantum Hardware}
Lastly, due to the limited number of PQC studies that perform executions on real quantum hardware, we evaluated the LQAS algorithm on two IBM devices, \textit{ibm{\textunderscore}fez} and \textit{ibm{\textunderscore}kingston}.
Due to the high number of device executions required by the LQAS approach, we only evaluated the base and final model produced through state-vector LQAS on IBM hardware.
Based on the performance of our state-vector simulations on the DDCC dataset, we chose this dataset as our proof-of-principle implementation.
As mentioned previously, this dataset is data-intensive in terms of the number of samples; therefore, we reduce the size of the training set from 80\% to 10\% of the total available, as training with 80\% of the data was computationally unfeasible based on preliminary experiments.
Additionally, the data are further divided into sets of four molecules per sample, where each molecule has 64 $t_{2}$-amplitudes or target values.
We chose to batch the data to reduce the number of circuit executions per primitive unified bloc (PUB), which has a maximum of 10 million evaluations per circuit, considering the number of observables, parameters, and shots.

The version of our code implemented using the \textsc{Qiskit} code mentioned in Subsection \ref{subsection:compdetails}, is verified using state-vector simulation for 500 iterations.
The base circuit has an $R^{2}$ of 0.836 and 0.830 and an MSE of 0.016 and 0.016, for the training and validation set, respectively.
Like with the larger dataset, there is an increase between the base and final LQAS model, where the training and validation set MSEs both reduce to 0.008, while the $R^{2}$s increase to 0.916 for the training set and 0.915 for the validation set.
Based on these promising state-vector results, we decided to run the base circuit on \textit{ibm{\textunderscore}fez} and the final circuit on \textit{ibm{\textunderscore}kingston}.
Overall, both experiments were run on the device for 500 iterations, which took a total of 14 days, 18 hours, 8 minutes, and 54 seconds of QPU time.
Additionally, the required QPU time corresponds to 72 days, 14 hours, 47 minutes, and 17 seconds of walltime.
As highlighted in Table~\ref{tab:ddcc_real_hw}, the deviation in performance between the base and final circuits is negligible. 

\begin{table}[h!]
    \centering
    \caption{Training and validation metrics for base vs. final models on real hardware.}
        \begin{tabular}{l l cccc}
            \toprule
            \multicolumn{1}{c}{\multirow{2}{*}{\parbox{2cm}{\centering Model}}} &
            \multicolumn{1}{c}{\multirow{2}{*}{\parbox{2cm}{\centering Device}}} &
            \multicolumn{2}{c}{Train} & \multicolumn{2}{c}{Validation} \\
            \cmidrule(lr){3-4} \cmidrule(lr){5-6}
             & & MSE & $R^2$ & MSE & $R^2$ \\
            \midrule
            Base  & \textit{ibm\_fez}      & 0.10741 & -0.12604 & 0.10837 & -0.13482 \\
            Final & \textit{ibm\_kingston} & 0.10327 & -0.08267 & 0.10548 & -0.10462 \\
            \bottomrule
        \end{tabular}
    \label{tab:ddcc_real_hw}
\end{table}



\section{Discussion}\label{section:discussion}
In this work, we introduced an evolution-inspired, heuristic algorithm based on a local, ansatz-level architecture search to iteratively improve PQCs for two synthetic and two quantum chemistry regression-based datasets.
The one- and two-dimensional quadratic function datasets, along with the DDCC dataset, highlighted the aptitude of the LQAS method using state-vector simulation.
One possible explanation for the performance of these three datasets is their relationship with the target value. 
In all three cases, the input (feature) values are directly related to the output (target) values.
For example, the top five features that are mapped to the quantum circuits for the DDCC dataset are used to initialize the optimized $t_{2}$-amplitudes (target values).
Despite these three datasets benefiting from a small number of features and qubits, the performance of the BSE49 dataset could be improved through wider quantum circuits containing more features.

The performance of the real quantum devices versus state-vector simulation highlights several difficulties regarding the evaluation of PQCs on real quantum hardware. 
One potential drawback of current quantum hardware, as highlighted by both the QPU time ($\sim14$ days) and walltime ($\sim72$ days), is that datasets that are high-dimensional regarding the number of samples are extremely costly. Another drawback is that even when circuit optimization and error mitigation is applied, model performance is lacking with respect to state-vector simulations.
Despite these potential drawbacks, we believe executing our algorithms on real quantum devices helps the field determine the gap between theory and experimentation.

\section{Conclusions}\label{section:Conclusions}
To summarize, in this work, we introduced an evolution-inspired algorithm based on a local, ansatz-level architecture search to iteratively improve PQCs for two synthetic and two quantum chemistry regression-based datasets.
Starting with the one-dimensional noisy quadratic dataset, we demonstrated that the LQAS algorithm was capable of improving model performance.
For example, for HEA-1-1, the base R$^{2}$ was improved from -0.993 to 0.958 by the final LQAS iteration.
Similarly, the two-dimensional quadratic dataset displayed similar improvements, such as the mean R$^{2}$ improving from 0.415 for the base circuits to 0.835 for the last LQAS iteration.
Following our evaluation of the synthetic datasets, we analyzed the performance of the LQAS on real quantum chemistry datasets using both state-vector simulation and real quantum devices.
Our state-vector results for both datasets illustrated that the DDCC dataset is capable of being modeled using PQCs with a few qubits, whereas wider circuits, more features, and more model parameters are required to adequately model the BSE49 dataset.
Additionally, our experiments on the real quantum devices highlight the current limitations of the current generation of hardware for PQC-based QML tasks.
Lastly, our data and source code are available in the following repository:  \href{https://github.com/MSRG/QMLpredict}{https://github.com/MSRG/QMLpredict}.

\begin{acks}
We acknowledge the Government of Canada’s New Frontiers in Research Fund (NFRF), for grant NFRFE-2022-00226, and the Quantum Software Consortium (QSC), financed under grant \#ALLRP587590-23 from the National Sciences and Engineering Research Council of Canada (NSERC) Alliance Consortia Quantum Grants.
This research was enabled in part by computational support provided by IBM Quantum via the Quantum Software Consortium and PINQ2, along with access to classical resources through the Digital Research Alliance of Canada.
Alexander Lao would like to acknowledge the Engineering Science Research Opportunities Program (ESROP) at the University of Toronto for funding.
Authors thank Anton Sugolov for his initial exploration of this work which was funded through the Natural Sciences and Engineering Research Council of Canada Undergraduate Student Research Awards (NSERC-USRA).
Additionally, Viki Kumar Prasad would like to acknowledge the Data Sciences Institute (DSI) for its support via the DSI Postdoctoral Fellowship throughout this project.

\end{acks}



\section*{Appendix}
\renewcommand{\thesection}{\Alph{section}}
\setcounter{table}{0}
\renewcommand{\tablename}{Table}
\renewcommand{\thetable}{\Alph{section}.\arabic{table}}

\setcounter{figure}{0}
\renewcommand{\figurename}{Figure}
\renewcommand{\thefigure}{\Alph{figure}}

\setcounter{section}{0}
\renewcommand{\thesection}{\Alph{section}}

\numberwithin{equation}{section}             
\renewcommand{\theequation}{\thesection.\arabic{equation}} 

\section{One- and Two-Dimensional Noisy Quadratic Function Model Performance}
\label{results:noisy-quadratic}

\begin{table}[H]
    \centering
    \caption{Performance metrics (MSE and $R^2$) for HEA ans\"{a}tze across base and LQAS iterations for the noisy one-dimensional quadratic function.}
    \label{tab:noisy-1Dquadratic}
    \begin{tabular}{lcccccccc}
        \hline
        \textbf{Ansatz} & \multicolumn{2}{c}{\textbf{Base}} & \multicolumn{2}{c}{\textbf{Iteration 1}} & \multicolumn{2}{c}{\textbf{Iteration 2}} & \multicolumn{2}{c}{\textbf{Iteration 3}} \\
         & MSE & $R^2$ & MSE & $R^2$ & MSE & $R^2$ & MSE & $R^2$ \\
        \hline
        HEA-1-1 & 2.534 & -0.993 & 0.062 & 0.951 & 0.070 & 0.945 & 0.054 & 0.958 \\
        HEA-1-2 & 0.125 & 0.902 & 0.010 & 0.992 & 0.012 & 0.991 & 0.008 & 0.993 \\
        HEA-1-3 & 0.042 & 0.967 & 0.009 & 0.993 & 0.006 & 0.995 & 0.006 & 0.996 \\
        HEA-2-1 & 0.041 & 0.968 & 0.024 & 0.981 & 0.014 & 0.989 & 0.014 & 0.989 \\
        HEA-2-2 & 0.015 & 0.989 & 0.007 & 0.995 & 0.005 & 0.996 & 0.007 & 0.995 \\
        HEA-2-3 & 0.011 & 0.992 & 0.006 & 0.996 & 0.007 & 0.995 & 0.005 & 0.996 \\
        HEA-3-1 & 0.044 & 0.965 & 0.012 & 0.991 & 0.011 & 0.991 & 0.011 & 0.992 \\
        HEA-3-2 & 0.046 & 0.964 & 0.003 & 0.998 & 0.003 & 0.998 & 0.006 & 0.995 \\
        HEA-3-3 & 0.054 & 0.957 & 0.007 & 0.995 & 0.006 & 0.995 & 0.004 & 0.997 \\
        \hline
    \end{tabular}
\end{table}

\begin{table}[H]
\centering
\caption{MSE and $R^2$ values across the base and LQAS iterations for different HEA ans\"{a}tze for the two-dimensional noisy quadratic dataset.}
\label{tab:noisy-2Dquadratic}
\begin{tabular}{lcccccccc}
\hline
\textbf{Ansatz} & \multicolumn{2}{c}{\textbf{Base}} & \multicolumn{2}{c}{\textbf{Iteration 1}} & \multicolumn{2}{c}{\textbf{Iteration 2}} & \multicolumn{2}{c}{\textbf{Iteration 3}} \\
 & MSE & $R^2$ & MSE & $R^2$ & MSE & $R^2$ & MSE & $R^2$ \\
\hline
HEA-1-1 & 0.204 & -0.520 & 0.162 & -0.207 & 0.161 & -0.198 & 0.116 & 0.137 \\
HEA-1-2 & 0.152 & -0.132 & 0.020 & 0.854 & 0.013 & 0.905 & 0.008 & 0.939 \\
HEA-1-3 & 0.015 & 0.888 & 0.007 & 0.951 & 0.006 & 0.956 & 0.007 & 0.951 \\
HEA-1-4 & 0.068 & 0.493 & 0.005 & 0.960 & 0.012 & 0.911 & 0.006 & 0.956 \\
HEA-2-2 & 0.045 & 0.667 & 0.008 & 0.943 & 0.010 & 0.926 & 0.006 & 0.954 \\
HEA-3-2 & 0.036 & 0.732 & 0.006 & 0.957 & 0.007 & 0.949 & 0.006 & 0.958 \\
HEA-4-2 & 0.030 & 0.778 & 0.008 & 0.939 & 0.009 & 0.932 & 0.006 & 0.953 \\
\hline
\end{tabular}
\end{table}

\section{DDCC Dataset Explanation}\label{section:dataset_ddcc}
\setcounter{table}{0}

To understand the DDCC method, as described in~\citet{Townsend2019-gx}, we briefly discuss the CC wave function and related expressions, where the wave function takes the following form,

\begin{equation}
\ket{\Psi_{\text{CC}}} = \exp \left( \hat{T} \right) \ket{\Psi_{0}}
\label{eq:cc_wfn}
\end{equation}

\noindent where $\hat{T}$ is the cluster operator and $\ket{\Psi_{0}}$ is the reference (Hartree-Fock) wave function.
In CCSD, the $\hat{T}$ operator truncated to only include single ($\hat{T}_{1}$) and double ($\hat{T}_{2}$) excitations. 
Following the evaluation of the CC equations, the CCSD correlation energy can then be formulated as such, 

\begin{equation}
	E^{\text{CCSD}}_{\text{corr}} = \sum_{\substack{a<b \\ i<j}} \mel{ij}{}{ab} t^{ab}_{ij} + \sum_{\substack{a<b \\ i<j}} \mel{ij}{}{ab} t^{a}_{i} t^{b}_{j}
	\label{eq:cc_corr}
\end{equation}
\noindent where $i$ and $j$ denote occupied orbitals, $a$ and $b$ denote virtual orbitals, $t^{ab}_{ij}$ correspond to two-electron excitation amplitudes ($t_{2}$-amplitudes), $t^{a}_{i}$ and $t^{b}_{j}$ correspond to one-electron excitation amplitudes ($t_{1}$-amplitudes), and $\mel{ij}{}{ab}$ are two-electron integrals.

The objective of this ML-based scheme is to learn the CCSD $t_{2}$-amplitudes using features generated from Hartree-Fock (HF) and M{\o}ller-Plesset second-order perturbation theory (MP2).
This task is straightforward since, the CCSD $t_{2}$-amplitudes are initialized using MP2 $t_{2}$-amplitudes, defined as,
\begin{equation}
	t^{ab}_{ij(\text{MP2})} = \frac{\mel{ij}{}{ab}}{\varepsilon_{i}+\varepsilon_{j}-\varepsilon_{a}-\varepsilon_{b}}
	\label{eq:MP2_t2}
\end{equation}
where $\varepsilon_{i}$ and $\varepsilon_{j}$ denote the orbital energies of the occupied orbitals $i$ and $j$, while the virtual orbitals $a$ and $b$ are denoted by $\varepsilon_{a}$ and $\varepsilon_{b}$.
The MP2 $t_{2}$-amplitudes are included in the DDCC dataset in addition to the numerator ($\mel{ij}{}{ab}$), denominator ($\varepsilon_{i}+\varepsilon_{j}-\varepsilon_{a}-\varepsilon_{b}$), a binary feature to denote whether the excitation goes to the same virtual orbital, and the orbital energies ($\varepsilon_{i},\varepsilon_{j},\varepsilon_{a},\varepsilon_{b}$).
Additional terms related to the individual contributions to the orbital energies, such as the one-electron Hamiltonian ($h$), Coulomb matrix ($J$), exchange matrix $K$, and Coulomb and exchange integrals ($J^{i}_{a}, J^{j}_{b}, K^{a}_{i}, K^{b}_{j}$), are included.
The original DDCC dataset consists of 30 features for each CCSD $t_{2}$-amplitude due to the addition of features that denote the sign and magnitude of the MP2 $t_{2}$-amplitudes.

\section{BSE49 Model Performance}\label{section:bse49modelperformance}
\setcounter{table}{0}

\begin{table}[H]
    \centering
    \caption{BSE49 training and validation metrics over five iterations of the LQAS algorithm.}
    \begin{tabular}{lcccc}
        \toprule
        \multirow{2}{*}{\textbf{Iteration}} & \multicolumn{2}{c}{Training Set} & \multicolumn{2}{c}{Validation Set} \\
        \cmidrule(lr){2-3} \cmidrule(lr){4-5}
        & Loss & $R^2$ & Loss & $R^2$ \\
        \midrule
        Base    & 0.148 & 0.313 & 0.072 & 0.348 \\
        Iter. 1 & 0.118 & 0.425 & 0.060 & 0.456 \\
        Iter. 2 & 0.189 & 0.425 & 0.057 & 0.484 \\
        Iter. 3 & 0.169 & 0.441 & 0.056 & 0.487 \\
        Iter. 4 & 0.203 & 0.441 & 0.056 & 0.490 \\
        Iter. 5 & 0.150 & 0.439 & 0.055 & 0.503 \\
        Iter. 6 & 0.167 & 0.443 & 0.056 & 0.498 \\
        Iter. 7 & 0.214 & 0.406 & 0.056 & 0.489 \\
        Iter. 8 & 0.188 & 0.439 & 0.057 & 0.482 \\
        Iter. 9 & 0.033 & 0.440 & 0.055 & 0.502 \\
        \bottomrule
    \end{tabular}
    \label{tab:training_metrics_bse16}
\end{table}

\section{DDCC Model Performance}\label{section:ddccmodelperformance}
\setcounter{table}{0}

\begin{table}[H]
    \centering
    \caption{Model performance on the DDCC dataset across five iterations of the LQAS algorithm.}
    \begin{tabular}{lcccc}
        \toprule
        Iteration & \multicolumn{2}{c}{Train} & \multicolumn{2}{c}{Validation} \\
        \cmidrule(lr){2-3} \cmidrule(lr){4-5}
        & Loss & $R^2$ & Loss & $R^2$ \\
        \midrule
        Base    & 0.356 & 0.946 & 0.0047 & 0.948 \\
        Iter. 1 & 0.236 & 0.986 & 0.0012 & 0.987 \\
        Iter. 2 & 0.062 & 0.991 & 0.0008 & 0.991 \\
        Iter. 3 & 0.177 & 0.989 & 0.00095 & 0.990 \\
        Iter. 4 & 0.108 & 0.991 & 0.0008 & 0.991 \\
        Iter. 5 & 0.133 & 0.993 & 0.0006 & 0.993 \\
        \bottomrule
    \end{tabular}
    \label{tab:ddccmodel}
\end{table}

\bibliographystyle{ACM-Reference-Format}
\bibliography{bib}


\begin{thebibliography}{73}


\ifx \showCODEN    \undefined \def \showCODEN     #1{\unskip}     \fi
\ifx \showISBNx    \undefined \def \showISBNx     #1{\unskip}     \fi
\ifx \showISBNxiii \undefined \def \showISBNxiii  #1{\unskip}     \fi
\ifx \showISSN     \undefined \def \showISSN      #1{\unskip}     \fi
\ifx \showLCCN     \undefined \def \showLCCN      #1{\unskip}     \fi
\ifx \shownote     \undefined \def \shownote      #1{#1}          \fi
\ifx \showarticletitle \undefined \def \showarticletitle #1{#1}   \fi
\ifx \showURL      \undefined \def \showURL       {\relax}        \fi
\providecommand\bibfield[2]{#2}
\providecommand\bibinfo[2]{#2}
\providecommand\natexlab[1]{#1}
\providecommand\showeprint[2][]{arXiv:#2}

\bibitem[mol({[n.\,d.]})]%
        {molssiMolSSIMLDatasets}
 \bibinfo{year}{[n.\,d.]}\natexlab{}.
\newblock \bibinfo{title}{{M}ol{S}{S}{I} {M}{L}-{D}atasets ---
  ml-datasets.molssi.org}.
\newblock \bibinfo{howpublished}{\url{https://ml-datasets.molssi.org/}}.
\newblock
\newblock
\shownote{[Accessed 28-01-2026]}.


\bibitem[qua({[n.\,d.]})]%
        {quantummachineQuantumMachineorgDatasets}
 \bibinfo{year}{[n.\,d.]}\natexlab{}.
\newblock \bibinfo{title}{{Q}uantum-{M}achine.org: {D}atasets ---
  quantum-machine.org}.
\newblock \bibinfo{howpublished}{\url{https://quantum-machine.org/datasets/}}.
\newblock
\newblock
\shownote{[Accessed 28-01-2026]}.


\bibitem[Altares-López et~al\mbox{.}(2021)]%
        {Altares-Lopez2021-yc}
\bibfield{author}{\bibinfo{person}{Sergio Altares-López},
  \bibinfo{person}{Angela Ribeiro}, {and} \bibinfo{person}{Juan~José
  García-Ripoll}.} \bibinfo{year}{2021}\natexlab{}.
\newblock \showarticletitle{Automatic design of quantum feature maps}.
\newblock \bibinfo{journal}{\emph{Quantum Sci. Technol.}} \bibinfo{volume}{6},
  \bibinfo{number}{4} (\bibinfo{date}{Aug.} \bibinfo{year}{2021}),
  \bibinfo{pages}{045015}.
\newblock


\bibitem[Benedetti et~al\mbox{.}(2019)]%
        {Benedetti2019-pf}
\bibfield{author}{\bibinfo{person}{Marcello Benedetti}, \bibinfo{person}{Erika
  Lloyd}, \bibinfo{person}{Stefan Sack}, {and} \bibinfo{person}{Mattia
  Fiorentini}.} \bibinfo{year}{2019}\natexlab{}.
\newblock \showarticletitle{Parameterized quantum circuits as machine learning
  models}.
\newblock \bibinfo{journal}{\emph{Quantum Sci. Technol.}} \bibinfo{volume}{4},
  \bibinfo{number}{4} (\bibinfo{date}{Nov.} \bibinfo{year}{2019}),
  \bibinfo{pages}{043001}.
\newblock


\bibitem[Bergholm et~al\mbox{.}(2018)]%
        {Bergholm2018-ex}
\bibfield{author}{\bibinfo{person}{Ville Bergholm}, \bibinfo{person}{Josh
  Izaac}, \bibinfo{person}{Maria Schuld}, \bibinfo{person}{Christian Gogolin},
  \bibinfo{person}{Shahnawaz Ahmed}, \bibinfo{person}{Vishnu Ajith},
  \bibinfo{person}{M~Sohaib Alam}, \bibinfo{person}{Guillermo Alonso-Linaje},
  \bibinfo{person}{B AkashNarayanan}, \bibinfo{person}{Ali Asadi},
  \bibinfo{person}{Juan~Miguel Arrazola}, \bibinfo{person}{Utkarsh Azad},
  \bibinfo{person}{Sam Banning}, \bibinfo{person}{Carsten Blank},
  \bibinfo{person}{Thomas~R Bromley}, \bibinfo{person}{Benjamin~A Cordier},
  \bibinfo{person}{Jack Ceroni}, \bibinfo{person}{Alain Delgado},
  \bibinfo{person}{Olivia Di~Matteo}, \bibinfo{person}{Amintor Dusko},
  \bibinfo{person}{Tanya Garg}, \bibinfo{person}{Diego Guala},
  \bibinfo{person}{Anthony Hayes}, \bibinfo{person}{Ryan Hill},
  \bibinfo{person}{Aroosa Ijaz}, \bibinfo{person}{Theodor Isacsson},
  \bibinfo{person}{David Ittah}, \bibinfo{person}{Soran Jahangiri},
  \bibinfo{person}{Prateek Jain}, \bibinfo{person}{Edward Jiang},
  \bibinfo{person}{Ankit Khandelwal}, \bibinfo{person}{Korbinian Kottmann},
  \bibinfo{person}{Robert~A Lang}, \bibinfo{person}{Christina Lee},
  \bibinfo{person}{Thomas Loke}, \bibinfo{person}{Angus Lowe},
  \bibinfo{person}{Keri McKiernan}, \bibinfo{person}{Johannes~Jakob Meyer},
  \bibinfo{person}{J~A Montañez-Barrera}, \bibinfo{person}{Romain Moyard},
  \bibinfo{person}{Zeyue Niu}, \bibinfo{person}{Lee~James O'Riordan},
  \bibinfo{person}{Steven Oud}, \bibinfo{person}{Ashish Panigrahi},
  \bibinfo{person}{Chae-Yeun Park}, \bibinfo{person}{Daniel Polatajko},
  \bibinfo{person}{Nicolás Quesada}, \bibinfo{person}{Chase Roberts},
  \bibinfo{person}{Nahum Sá}, \bibinfo{person}{Isidor Schoch},
  \bibinfo{person}{Borun Shi}, \bibinfo{person}{Shuli Shu},
  \bibinfo{person}{Sukin Sim}, \bibinfo{person}{Arshpreet Singh},
  \bibinfo{person}{Ingrid Strandberg}, \bibinfo{person}{Jay Soni},
  \bibinfo{person}{Antal Száva}, \bibinfo{person}{Slimane Thabet},
  \bibinfo{person}{Rodrigo~A Vargas-Hernández}, \bibinfo{person}{Trevor
  Vincent}, \bibinfo{person}{Nicola Vitucci}, \bibinfo{person}{Maurice Weber},
  \bibinfo{person}{David Wierichs}, \bibinfo{person}{Roeland Wiersema},
  \bibinfo{person}{Moritz Willmann}, \bibinfo{person}{Vincent Wong},
  \bibinfo{person}{Shaoming Zhang}, {and} \bibinfo{person}{Nathan Killoran}.}
  \bibinfo{year}{2018}\natexlab{}.
\newblock \showarticletitle{{PennyLane}: Automatic differentiation of hybrid
  quantum-classical computations}.
\newblock \bibinfo{journal}{\emph{arXiv [quant-ph]}} (\bibinfo{date}{Nov.}
  \bibinfo{year}{2018}).
\newblock


\bibitem[Bhatia et~al\mbox{.}(2023)]%
        {bhatia_quantum_2023}
\bibfield{author}{\bibinfo{person}{Amandeep~Singh Bhatia},
  \bibinfo{person}{Mandeep~Kaur Saggi}, {and} \bibinfo{person}{Sabre Kais}.}
  \bibinfo{year}{2023}\natexlab{}.
\newblock \showarticletitle{Quantum {Machine} {Learning} {Predicting}
  {ADME}-{Tox} {Properties} in {Drug} {Discovery}}.
\newblock \bibinfo{journal}{\emph{J. Chem. Inf. Model.}} \bibinfo{volume}{63},
  \bibinfo{number}{21} (\bibinfo{date}{Nov.} \bibinfo{year}{2023}),
  \bibinfo{pages}{6476--6486}.
\newblock
\showISSN{1549-9596}
\href{https://doi.org/10.1021/acs.jcim.3c01079}{doi:\nolinkurl{10.1021/acs.jcim.3c01079}}


\bibitem[Biamonte et~al\mbox{.}(2017)]%
        {Biamonte2017-hp}
\bibfield{author}{\bibinfo{person}{Jacob Biamonte}, \bibinfo{person}{Peter
  Wittek}, \bibinfo{person}{Nicola Pancotti}, \bibinfo{person}{Patrick
  Rebentrost}, \bibinfo{person}{Nathan Wiebe}, {and} \bibinfo{person}{Seth
  Lloyd}.} \bibinfo{year}{2017}\natexlab{}.
\newblock \showarticletitle{Quantum machine learning}.
\newblock \bibinfo{journal}{\emph{Nature}} \bibinfo{volume}{549},
  \bibinfo{number}{7671} (\bibinfo{date}{Sept.} \bibinfo{year}{2017}),
  \bibinfo{pages}{195--202}.
\newblock


\bibitem[Bowles et~al\mbox{.}(2023)]%
        {Bowles2023-cl}
\bibfield{author}{\bibinfo{person}{Joseph Bowles}, \bibinfo{person}{Victoria~J
  Wright}, \bibinfo{person}{M{\'a}t{\'e} Farkas}, \bibinfo{person}{Nathan
  Killoran}, {and} \bibinfo{person}{Maria Schuld}.}
  \bibinfo{year}{2023}\natexlab{}.
\newblock \showarticletitle{Contextuality and inductive bias in quantum machine
  learning}.
\newblock \bibinfo{journal}{\emph{arXiv preprint arXiv:2302.01365}}
  (\bibinfo{year}{2023}).
\newblock


\bibitem[Butler et~al\mbox{.}(2018)]%
        {Butler2018-io}
\bibfield{author}{\bibinfo{person}{Keith~T Butler}, \bibinfo{person}{Daniel~W
  Davies}, \bibinfo{person}{Hugh Cartwright}, \bibinfo{person}{Olexandr
  Isayev}, {and} \bibinfo{person}{Aron Walsh}.}
  \bibinfo{year}{2018}\natexlab{}.
\newblock \showarticletitle{Machine learning for molecular and materials
  science}.
\newblock \bibinfo{journal}{\emph{Nature}} \bibinfo{volume}{559},
  \bibinfo{number}{7715} (\bibinfo{date}{July} \bibinfo{year}{2018}),
  \bibinfo{pages}{547--555}.
\newblock


\bibitem[Cao et~al\mbox{.}(2019)]%
        {Cao2019-lh}
\bibfield{author}{\bibinfo{person}{Yudong Cao}, \bibinfo{person}{Jonathan
  Romero}, \bibinfo{person}{Jonathan~P Olson}, \bibinfo{person}{Matthias
  Degroote}, \bibinfo{person}{Peter~D Johnson}, \bibinfo{person}{Mária
  Kieferová}, \bibinfo{person}{Ian~D Kivlichan}, \bibinfo{person}{Tim Menke},
  \bibinfo{person}{Borja Peropadre}, \bibinfo{person}{Nicolas P~D Sawaya},
  \bibinfo{person}{Sukin Sim}, \bibinfo{person}{Libor Veis}, {and}
  \bibinfo{person}{Alán Aspuru-Guzik}.} \bibinfo{year}{2019}\natexlab{}.
\newblock \showarticletitle{Quantum chemistry in the age of quantum computing}.
\newblock \bibinfo{journal}{\emph{Chem. Rev.}} \bibinfo{volume}{119},
  \bibinfo{number}{19} (\bibinfo{date}{Oct.} \bibinfo{year}{2019}),
  \bibinfo{pages}{10856--10915}.
\newblock


\bibitem[Caro et~al\mbox{.}(2022)]%
        {Caro2022-uf}
\bibfield{author}{\bibinfo{person}{Matthias~C Caro}, \bibinfo{person}{Hsin-Yuan
  Huang}, \bibinfo{person}{M Cerezo}, \bibinfo{person}{Kunal Sharma},
  \bibinfo{person}{Andrew Sornborger}, \bibinfo{person}{Lukasz Cincio}, {and}
  \bibinfo{person}{Patrick~J Coles}.} \bibinfo{year}{2022}\natexlab{}.
\newblock \showarticletitle{Generalization in quantum machine learning from few
  training data}.
\newblock \bibinfo{journal}{\emph{Nat. Commun.}} \bibinfo{volume}{13},
  \bibinfo{number}{1} (\bibinfo{year}{2022}), \bibinfo{pages}{4919}.
\newblock


\bibitem[Cerezo et~al\mbox{.}(2021)]%
        {Cerezo2021-fc}
\bibfield{author}{\bibinfo{person}{M Cerezo}, \bibinfo{person}{Andrew
  Arrasmith}, \bibinfo{person}{Ryan Babbush}, \bibinfo{person}{Simon~C
  Benjamin}, \bibinfo{person}{Suguru Endo}, \bibinfo{person}{Keisuke Fujii},
  \bibinfo{person}{Jarrod~R McClean}, \bibinfo{person}{Kosuke Mitarai},
  \bibinfo{person}{Xiao Yuan}, \bibinfo{person}{Lukasz Cincio}, {and}
  \bibinfo{person}{Patrick~J Coles}.} \bibinfo{year}{2021}\natexlab{}.
\newblock \showarticletitle{Variational quantum algorithms}.
\newblock \bibinfo{journal}{\emph{Nat. Rev. Phys.}} \bibinfo{volume}{3},
  \bibinfo{number}{9} (\bibinfo{date}{Aug.} \bibinfo{year}{2021}),
  \bibinfo{pages}{625–644}.
\newblock


\bibitem[Chivilikhin et~al\mbox{.}(2020)]%
        {Chivilikhin2020-ea}
\bibfield{author}{\bibinfo{person}{D Chivilikhin}, \bibinfo{person}{A Samarin},
  \bibinfo{person}{V Ulyantsev}, \bibinfo{person}{I Iorsh},
  \bibinfo{person}{A~R Oganov}, {and} \bibinfo{person}{O Kyriienko}.}
  \bibinfo{year}{2020}\natexlab{}.
\newblock \bibinfo{title}{{MoG}-{VQE}: Multiobjective genetic variational
  quantum eigensolver}.
\newblock


\bibitem[Deb et~al\mbox{.}(2002)]%
        {Deb2002-zx}
\bibfield{author}{\bibinfo{person}{K Deb}, \bibinfo{person}{A Pratap},
  \bibinfo{person}{S Agarwal}, {and} \bibinfo{person}{T Meyarivan}.}
  \bibinfo{year}{2002}\natexlab{}.
\newblock \showarticletitle{A fast and elitist multiobjective genetic
  algorithm: {NSGA}-{II}}.
\newblock \bibinfo{journal}{\emph{IEEE Trans. Evol. Comput.}}
  \bibinfo{volume}{6}, \bibinfo{number}{2} (\bibinfo{year}{2002}),
  \bibinfo{pages}{182--197}.
\newblock


\bibitem[Du et~al\mbox{.}(2022)]%
        {Du2022-zd}
\bibfield{author}{\bibinfo{person}{Yuxuan Du}, \bibinfo{person}{Tao Huang},
  \bibinfo{person}{Shan You}, \bibinfo{person}{Min-Hsiu Hsieh}, {and}
  \bibinfo{person}{Dacheng Tao}.} \bibinfo{year}{2022}\natexlab{}.
\newblock \showarticletitle{Quantum circuit architecture search for variational
  quantum algorithms}.
\newblock \bibinfo{journal}{\emph{npj Quantum Inf.}} \bibinfo{volume}{8},
  \bibinfo{number}{1} (\bibinfo{year}{2022}), \bibinfo{pages}{62}.
\newblock


\bibitem[García-Andrade et~al\mbox{.}(2023)]%
        {garcia-andrade_barrier_2023}
\bibfield{author}{\bibinfo{person}{Xabier García-Andrade},
  \bibinfo{person}{Pablo García~Tahoces}, \bibinfo{person}{Jesús
  Pérez-Ríos}, {and} \bibinfo{person}{Emilio Martínez~Núñez}.}
  \bibinfo{year}{2023}\natexlab{}.
\newblock \showarticletitle{Barrier {Height} {Prediction} by {Machine}
  {Learning} {Correction} of {Semiempirical} {Calculations}}.
\newblock \bibinfo{journal}{\emph{J. Phys. Chem. A}} \bibinfo{volume}{127},
  \bibinfo{number}{10} (\bibinfo{date}{March} \bibinfo{year}{2023}),
  \bibinfo{pages}{2274--2283}.
\newblock
\showISSN{1089-5639}
\href{https://doi.org/10.1021/acs.jpca.2c08340}{doi:\nolinkurl{10.1021/acs.jpca.2c08340}}


\bibitem[Goh et~al\mbox{.}(2017)]%
        {Goh2017-ay}
\bibfield{author}{\bibinfo{person}{Garrett~B Goh}, \bibinfo{person}{Nathan~O
  Hodas}, {and} \bibinfo{person}{Abhinav Vishnu}.}
  \bibinfo{year}{2017}\natexlab{}.
\newblock \showarticletitle{Deep learning for computational chemistry}.
\newblock \bibinfo{journal}{\emph{J. Comput. Chem.}} \bibinfo{volume}{38},
  \bibinfo{number}{16} (\bibinfo{date}{June} \bibinfo{year}{2017}),
  \bibinfo{pages}{1291--1307}.
\newblock


\bibitem[Hansen et~al\mbox{.}(2015)]%
        {Hansen2015-en}
\bibfield{author}{\bibinfo{person}{Katja Hansen}, \bibinfo{person}{Franziska
  Biegler}, \bibinfo{person}{Raghunathan Ramakrishnan}, \bibinfo{person}{Wiktor
  Pronobis}, \bibinfo{person}{O~Anatole von Lilienfeld},
  \bibinfo{person}{Klaus-Robert Müller}, {and} \bibinfo{person}{Alexandre
  Tkatchenko}.} \bibinfo{year}{2015}\natexlab{}.
\newblock \showarticletitle{Machine learning predictions of molecular
  properties: Accurate many-body potentials and nonlocality in chemical space}.
\newblock \bibinfo{journal}{\emph{J. Phys. Chem. Lett.}} \bibinfo{volume}{6},
  \bibinfo{number}{12} (\bibinfo{date}{June} \bibinfo{year}{2015}),
  \bibinfo{pages}{2326--2331}.
\newblock


\bibitem[Harrow and Montanaro(2017)]%
        {Harrow2017-eq}
\bibfield{author}{\bibinfo{person}{Aram~W Harrow} {and} \bibinfo{person}{Ashley
  Montanaro}.} \bibinfo{year}{2017}\natexlab{}.
\newblock \showarticletitle{Quantum computational supremacy}.
\newblock \bibinfo{journal}{\emph{Nature}} \bibinfo{volume}{549},
  \bibinfo{number}{7671} (\bibinfo{date}{Sept.} \bibinfo{year}{2017}),
  \bibinfo{pages}{203–209}.
\newblock


\bibitem[Hatakeyama-Sato et~al\mbox{.}(2023)]%
        {Hatakeyama-Sato2023-lr}
\bibfield{author}{\bibinfo{person}{Kan Hatakeyama-Sato},
  \bibinfo{person}{Yasuhiko Igarashi}, \bibinfo{person}{Takahiro Kashikawa},
  \bibinfo{person}{Koichi Kimura}, {and} \bibinfo{person}{Kenichi Oyaizu}.}
  \bibinfo{year}{2023}\natexlab{}.
\newblock \showarticletitle{Quantum circuit learning as a potential algorithm
  to predict experimental chemical properties}.
\newblock \bibinfo{journal}{\emph{Digit. Discov.}} \bibinfo{volume}{2},
  \bibinfo{number}{1} (\bibinfo{year}{2023}), \bibinfo{pages}{165--176}.
\newblock


\bibitem[Hehre et~al\mbox{.}(1970)]%
        {hehre_selfconsistent_1970}
\bibfield{author}{\bibinfo{person}{W.~J. Hehre}, \bibinfo{person}{R.
  Ditchfield}, \bibinfo{person}{R.~F. Stewart}, {and} \bibinfo{person}{J.~A.
  Pople}.} \bibinfo{year}{1970}\natexlab{}.
\newblock \showarticletitle{Self‐{Consistent} {Molecular} {Orbital}
  {Methods}. {IV}. {Use} of {Gaussian} {Expansions} of {Slater}‐{Type}
  {Orbitals}. {Extension} to {Second}‐{Row} {Molecules}}.
\newblock \bibinfo{journal}{\emph{J. Chem. Phys.}} \bibinfo{volume}{52},
  \bibinfo{number}{5} (\bibinfo{date}{March} \bibinfo{year}{1970}),
  \bibinfo{pages}{2769--2773}.
\newblock
\showISSN{0021-9606}
\href{https://doi.org/10.1063/1.1673374}{doi:\nolinkurl{10.1063/1.1673374}}


\bibitem[Heid et~al\mbox{.}(2023)]%
        {Heid2023-cu}
\bibfield{author}{\bibinfo{person}{Esther Heid}, \bibinfo{person}{Kevin~P
  Greenman}, \bibinfo{person}{Yunsie Chung}, \bibinfo{person}{Shih-Cheng Li},
  \bibinfo{person}{David~E Graff}, \bibinfo{person}{Florence~H Vermeire},
  \bibinfo{person}{Haoyang Wu}, \bibinfo{person}{William~H Green}, {and}
  \bibinfo{person}{Charles~J McGill}.} \bibinfo{year}{2023}\natexlab{}.
\newblock \showarticletitle{Chemprop: A Machine Learning Package for Chemical
  Property Prediction}.
\newblock \bibinfo{journal}{\emph{J. Chem. Inf. Model.}} \bibinfo{volume}{64},
  \bibinfo{number}{1} (\bibinfo{date}{Dec.} \bibinfo{year}{2023}),
  \bibinfo{pages}{9–17}.
\newblock


\bibitem[Holmes et~al\mbox{.}(2022)]%
        {Holmes2022-gc}
\bibfield{author}{\bibinfo{person}{Zoë Holmes}, \bibinfo{person}{Kunal
  Sharma}, \bibinfo{person}{M Cerezo}, {and} \bibinfo{person}{Patrick~J
  Coles}.} \bibinfo{year}{2022}\natexlab{}.
\newblock \showarticletitle{Connecting Ansatz Expressibility to Gradient
  Magnitudes and Barren Plateaus}.
\newblock \bibinfo{journal}{\emph{PRX quantum}}  \bibinfo{volume}{3}
  (\bibinfo{date}{Jan.} \bibinfo{year}{2022}), \bibinfo{pages}{010313}.
\newblock


\bibitem[Huang et~al\mbox{.}(2022)]%
        {Huang2022-as}
\bibfield{author}{\bibinfo{person}{Yuhan Huang}, \bibinfo{person}{Qingyu Li},
  \bibinfo{person}{Xiaokai Hou}, \bibinfo{person}{Rebing Wu},
  \bibinfo{person}{Man-Hong Yung}, \bibinfo{person}{Abolfazl Bayat}, {and}
  \bibinfo{person}{Xiaoting Wang}.} \bibinfo{year}{2022}\natexlab{}.
\newblock \showarticletitle{Robust resource-efficient quantum variational
  ansatz through an evolutionary algorithm}.
\newblock \bibinfo{journal}{\emph{Phys. Rev. A}} \bibinfo{volume}{105},
  \bibinfo{number}{5} (\bibinfo{year}{2022}), \bibinfo{pages}{052414}.
\newblock


\bibitem[Janet and Kulik(2020)]%
        {Janet2020-rl}
\bibfield{author}{\bibinfo{person}{Jon~Paul Janet} {and}
  \bibinfo{person}{Heather~J. Kulik}.} \bibinfo{year}{2020}\natexlab{}.
\newblock \bibinfo{booktitle}{\emph{Machine Learning in Chemistry}}.
\newblock \bibinfo{publisher}{American Chemical Society},
  \bibinfo{address}{Washington, DC, USA}.
\newblock
\href{https://doi.org/10.1021/acs.infocus.7e4001}{doi:\nolinkurl{10.1021/acs.infocus.7e4001}}


\bibitem[Javadi-Abhari et~al\mbox{.}(2024)]%
        {Javadi-Abhari2024-nw}
\bibfield{author}{\bibinfo{person}{Ali Javadi-Abhari}, \bibinfo{person}{Matthew
  Treinish}, \bibinfo{person}{Kevin Krsulich}, \bibinfo{person}{Christopher~J
  Wood}, \bibinfo{person}{Jake Lishman}, \bibinfo{person}{Julien Gacon},
  \bibinfo{person}{Simon Martiel}, \bibinfo{person}{Paul~D Nation},
  \bibinfo{person}{Lev~S Bishop}, \bibinfo{person}{Andrew~W Cross},
  \bibinfo{person}{Blake~R Johnson}, {and} \bibinfo{person}{Jay~M Gambetta}.}
  \bibinfo{year}{2024}\natexlab{}.
\newblock \showarticletitle{Quantum computing with Qiskit}.
\newblock \bibinfo{journal}{\emph{arXiv [quant-ph]}} (\bibinfo{date}{May}
  \bibinfo{year}{2024}).
\newblock


\bibitem[Jin and Merz(2025)]%
        {jin_integrating_2025}
\bibfield{author}{\bibinfo{person}{Hongni Jin} {and} \bibinfo{person}{Kenneth
  M.~Jr Merz}.} \bibinfo{year}{2025}\natexlab{}.
\newblock \showarticletitle{Integrating {Machine} {Learning} and {Quantum}
  {Circuits} for {Proton} {Affinity} {Predictions}}.
\newblock \bibinfo{journal}{\emph{J. Chem. Theory Comput.}}
  (\bibinfo{date}{Feb.} \bibinfo{year}{2025}).
\newblock
\showISSN{1549-9618}
\href{https://doi.org/10.1021/acs.jctc.4c01609}{doi:\nolinkurl{10.1021/acs.jctc.4c01609}}


\bibitem[Jones et~al\mbox{.}(2025)]%
        {Jones2025-kt}
\bibfield{author}{\bibinfo{person}{Grier~M Jones}, \bibinfo{person}{Viki~Kumar
  Prasad}, \bibinfo{person}{Ulrich Fekl}, {and} \bibinfo{person}{Hans-Arno
  Jacobsen}.} \bibinfo{year}{2025}\natexlab{}.
\newblock \showarticletitle{Parametrized Quantum Circuit Learning for Quantum
  Chemical Applications}.
\newblock \bibinfo{journal}{\emph{arXiv [quant-ph]}} (\bibinfo{date}{July}
  \bibinfo{year}{2025}).
\newblock


\bibitem[Jones et~al\mbox{.}(2023a)]%
        {Jones2023-hz}
\bibfield{author}{\bibinfo{person}{Grier~M Jones}, \bibinfo{person}{P~D~Varuna
  S.~Pathirage}, {and} \bibinfo{person}{Konstantinos~D Vogiatzis}.}
  \bibinfo{year}{2023}\natexlab{a}.
\newblock \showarticletitle{Chapter 22 - {Data}-driven acceleration of
  coupled-cluster and perturbation theory methods}.
\newblock In \bibinfo{booktitle}{\emph{Quantum {Chemistry} in the {Age} of
  {Machine} {Learning}}}, \bibfield{editor}{\bibinfo{person}{Pavlo~O Dral}}
  (Ed.). \bibinfo{publisher}{Elsevier}, \bibinfo{pages}{509--529}.
\newblock


\bibitem[Jones et~al\mbox{.}(2023b)]%
        {Jones2023-kh}
\bibfield{author}{\bibinfo{person}{Grier~M Jones}, \bibinfo{person}{Brett~A
  Smith}, \bibinfo{person}{Justin~K Kirkland}, {and}
  \bibinfo{person}{Konstantinos~D Vogiatzis}.}
  \bibinfo{year}{2023}\natexlab{b}.
\newblock \showarticletitle{Data-driven ligand field exploration of
  Fe(iv)–oxo sites for {C–H} activation}.
\newblock \bibinfo{journal}{\emph{Inorg. Chem. Front.}} \bibinfo{volume}{10},
  \bibinfo{number}{4} (\bibinfo{year}{2023}), \bibinfo{pages}{1062--1075}.
\newblock


\bibitem[Kandala et~al\mbox{.}(2017)]%
        {kandala2017hardware}
\bibfield{author}{\bibinfo{person}{Abhinav Kandala}, \bibinfo{person}{Antonio
  Mezzacapo}, \bibinfo{person}{Kristan Temme}, \bibinfo{person}{Maika Takita},
  \bibinfo{person}{Markus Brink}, \bibinfo{person}{Jerry~M Chow}, {and}
  \bibinfo{person}{Jay~M Gambetta}.} \bibinfo{year}{2017}\natexlab{}.
\newblock \showarticletitle{Hardware-efficient variational quantum eigensolver
  for small molecules and quantum magnets}.
\newblock \bibinfo{journal}{\emph{Nature}} \bibinfo{volume}{549},
  \bibinfo{number}{7671} (\bibinfo{year}{2017}), \bibinfo{pages}{242--246}.
\newblock


\bibitem[Kingma and Ba(2014)]%
        {Kingma2014-tz}
\bibfield{author}{\bibinfo{person}{Diederik~P Kingma} {and}
  \bibinfo{person}{Jimmy Ba}.} \bibinfo{year}{2014}\natexlab{}.
\newblock \showarticletitle{Adam: A method for stochastic optimization}.
\newblock \bibinfo{journal}{\emph{arXiv [cs.LG]}} (\bibinfo{date}{Dec.}
  \bibinfo{year}{2014}).
\newblock


\bibitem[Kumar et~al\mbox{.}(2020)]%
        {Kumar2020-ub}
\bibfield{author}{\bibinfo{person}{Chejarla~Santosh Kumar},
  \bibinfo{person}{Movva Naga~Sumanth Choudary}, \bibinfo{person}{Vinay~Babu
  Bommineni}, \bibinfo{person}{Grandhi Tarun}, {and} \bibinfo{person}{T
  Anjali}.} \bibinfo{year}{2020}\natexlab{}.
\newblock \showarticletitle{Dimensionality Reduction based on {SHAP} Analysis:
  A Simple and Trustworthy Approach}. In \bibinfo{booktitle}{\emph{2020
  International Conference on Communication and Signal Processing (ICCSP)}}.
  \bibinfo{pages}{558--560}.
\newblock


\bibitem[Landrum et~al\mbox{.}(2023)]%
        {Landrum2023-ov}
\bibfield{author}{\bibinfo{person}{Greg Landrum}, \bibinfo{person}{Paolo
  Tosco}, \bibinfo{person}{Brian Kelley}, \bibinfo{person}{{Ric}},
  \bibinfo{person}{{Sriniker}}, \bibinfo{person}{David Cosgrove},
  \bibinfo{person}{{Gedeck}}, \bibinfo{person}{Riccardo Vianello},
  \bibinfo{person}{{NadineSchneider}}, \bibinfo{person}{Eisuke Kawashima},
  \bibinfo{person}{Dan N}, \bibinfo{person}{Gareth Jones},
  \bibinfo{person}{Andrew Dalke}, \bibinfo{person}{Brian Cole},
  \bibinfo{person}{Matt Swain}, \bibinfo{person}{Samo Turk},
  \bibinfo{person}{{AlexanderSavelyev}}, \bibinfo{person}{Alain Vaucher},
  \bibinfo{person}{Maciej Wójcikowski}, \bibinfo{person}{{Ichiru Take}},
  \bibinfo{person}{Daniel Probst}, \bibinfo{person}{Kazuya Ujihara},
  \bibinfo{person}{Vincent~F Scalfani}, \bibinfo{person}{Guillaume Godin},
  \bibinfo{person}{Axel Pahl}, \bibinfo{person}{{Francois Berenger}},
  \bibinfo{person}{{JLVarjo}}, \bibinfo{person}{Rachel Walker},
  \bibinfo{person}{{Jasondbiggs}}, {and} \bibinfo{person}{{Strets123}}.}
  \bibinfo{year}{2023}\natexlab{}.
\newblock \bibinfo{title}{rdkit/rdkit: {2023\_03\_1} ({Q1} 2023) Release}.
\newblock


\bibitem[Lu et~al\mbox{.}(2021)]%
        {Lu2021-mx}
\bibfield{author}{\bibinfo{person}{Zhide Lu}, \bibinfo{person}{Pei-Xin Shen},
  {and} \bibinfo{person}{Dong-Ling Deng}.} \bibinfo{year}{2021}\natexlab{}.
\newblock \showarticletitle{Markovian quantum neuroevolution for machine
  learning}.
\newblock \bibinfo{journal}{\emph{Phys. Rev. Appl.}} \bibinfo{volume}{16},
  \bibinfo{number}{4} (\bibinfo{year}{2021}), \bibinfo{pages}{044039}.
\newblock


\bibitem[Ma et~al\mbox{.}(2025)]%
        {Ma2025-lp}
\bibfield{author}{\bibinfo{person}{Quangong Ma}, \bibinfo{person}{Chaolong
  Hao}, \bibinfo{person}{Nianwen Si}, {and} \bibinfo{person}{Dan Qu}.}
  \bibinfo{year}{2025}\natexlab{}.
\newblock \showarticletitle{Quantum architecture search for optimizing quantum
  generators in quantum {GAN}}.
\newblock \bibinfo{journal}{\emph{Mach. Learn. Sci. Technol.}}
  \bibinfo{volume}{6}, \bibinfo{number}{3} (\bibinfo{date}{Sept.}
  \bibinfo{year}{2025}), \bibinfo{pages}{035061}.
\newblock


\bibitem[Martyniuk et~al\mbox{.}(2024)]%
        {Martyniuk2024-cd}
\bibfield{author}{\bibinfo{person}{Darya Martyniuk}, \bibinfo{person}{Johannes
  Jung}, {and} \bibinfo{person}{Adrian Paschke}.}
  \bibinfo{year}{2024}\natexlab{}.
\newblock \showarticletitle{Quantum architecture search: a survey}. In
  \bibinfo{booktitle}{\emph{2024 IEEE International Conference on Quantum
  Computing and Engineering (QCE)}}, Vol.~\bibinfo{volume}{1}. IEEE,
  \bibinfo{pages}{1695--1706}.
\newblock


\bibitem[Maćkiewicz and Ratajczak(1993)]%
        {Mackiewicz1993-fc}
\bibfield{author}{\bibinfo{person}{Andrzej Maćkiewicz} {and}
  \bibinfo{person}{Waldemar Ratajczak}.} \bibinfo{year}{1993}\natexlab{}.
\newblock \showarticletitle{Principal components analysis ({PCA})}.
\newblock \bibinfo{journal}{\emph{Comput. Geosci.}} \bibinfo{volume}{19},
  \bibinfo{number}{3} (\bibinfo{date}{March} \bibinfo{year}{1993}),
  \bibinfo{pages}{303–342}.
\newblock


\bibitem[McClean et~al\mbox{.}(2016)]%
        {McClean2015-nv}
\bibfield{author}{\bibinfo{person}{Jarrod~R McClean}, \bibinfo{person}{Jonathan
  Romero}, \bibinfo{person}{Ryan Babbush}, {and} \bibinfo{person}{Al{\'a}n
  Aspuru-Guzik}.} \bibinfo{year}{2016}\natexlab{}.
\newblock \showarticletitle{The theory of variational hybrid quantum-classical
  algorithms}.
\newblock \bibinfo{journal}{\emph{New J. Phys.}} \bibinfo{volume}{18},
  \bibinfo{number}{2} (\bibinfo{year}{2016}), \bibinfo{pages}{023023}.
\newblock


\bibitem[Montgomery et~al\mbox{.}(1999)]%
        {Montgomery1999-fr}
\bibfield{author}{\bibinfo{person}{J~A Montgomery, Jr}, \bibinfo{person}{M~J
  Frisch}, \bibinfo{person}{J~W Ochterski}, {and} \bibinfo{person}{G~A
  Petersson}.} \bibinfo{year}{1999}\natexlab{}.
\newblock \showarticletitle{A complete basis set model chemistry. {VI}. Use of
  density functional geometries and frequencies}.
\newblock \bibinfo{journal}{\emph{J. Chem. Phys.}} \bibinfo{volume}{110},
  \bibinfo{number}{6} (\bibinfo{date}{Feb.} \bibinfo{year}{1999}),
  \bibinfo{pages}{2822--2827}.
\newblock


\bibitem[Montgomery et~al\mbox{.}(2000)]%
        {Montgomery2000-cn}
\bibfield{author}{\bibinfo{person}{J~A Montgomery, Jr}, \bibinfo{person}{M~J
  Frisch}, \bibinfo{person}{J~W Ochterski}, {and} \bibinfo{person}{G~A
  Petersson}.} \bibinfo{year}{2000}\natexlab{}.
\newblock \showarticletitle{A complete basis set model chemistry. {VII}. Use of
  the minimum population localization method}.
\newblock \bibinfo{journal}{\emph{J. Chem. Phys.}} \bibinfo{volume}{112},
  \bibinfo{number}{15} (\bibinfo{date}{April} \bibinfo{year}{2000}),
  \bibinfo{pages}{6532--6542}.
\newblock


\bibitem[Morgan(1965)]%
        {Morgan1965-kv}
\bibfield{author}{\bibinfo{person}{H~L Morgan}.}
  \bibinfo{year}{1965}\natexlab{}.
\newblock \showarticletitle{The generation of a unique machine description for
  chemical structures-A technique developed at chemical abstracts service}.
\newblock \bibinfo{journal}{\emph{J. Chem. Doc.}} \bibinfo{volume}{5},
  \bibinfo{number}{2} (\bibinfo{date}{May} \bibinfo{year}{1965}),
  \bibinfo{pages}{107--113}.
\newblock


\bibitem[Nandy et~al\mbox{.}(2021)]%
        {Nandy2021-qg}
\bibfield{author}{\bibinfo{person}{Aditya Nandy}, \bibinfo{person}{Chenru
  Duan}, \bibinfo{person}{Michael~G Taylor}, \bibinfo{person}{Fang Liu},
  \bibinfo{person}{Adam~H Steeves}, {and} \bibinfo{person}{Heather~J Kulik}.}
  \bibinfo{year}{2021}\natexlab{}.
\newblock \showarticletitle{Computational discovery of transition-metal
  complexes: From high-throughput screening to machine learning}.
\newblock \bibinfo{journal}{\emph{Chem. Rev.}} \bibinfo{volume}{121},
  \bibinfo{number}{16} (\bibinfo{date}{Aug.} \bibinfo{year}{2021}),
  \bibinfo{pages}{9927--10000}.
\newblock


\bibitem[Ostaszewski et~al\mbox{.}(2021)]%
        {Ostaszewski2021-dd}
\bibfield{author}{\bibinfo{person}{Mateusz Ostaszewski}, \bibinfo{person}{Lea~M
  Trenkwalder}, \bibinfo{person}{Wojciech Masarczyk}, \bibinfo{person}{Eleanor
  Scerri}, {and} \bibinfo{person}{Vedran Dunjko}.}
  \bibinfo{year}{2021}\natexlab{}.
\newblock \showarticletitle{Reinforcement learning for optimization of
  variational quantum circuit architectures}.
\newblock \bibinfo{journal}{\emph{Adv. Neural Inf. Process. Syst.}}
  \bibinfo{volume}{34} (\bibinfo{year}{2021}), \bibinfo{pages}{18182--18194}.
\newblock


\bibitem[Parrish et~al\mbox{.}(2017)]%
        {parrish_psi4_2017}
\bibfield{author}{\bibinfo{person}{Robert~M. Parrish}, \bibinfo{person}{Lori~A.
  Burns}, \bibinfo{person}{Daniel G.~A. Smith}, \bibinfo{person}{Andrew~C.
  Simmonett}, \bibinfo{person}{A.~Eugene~III DePrince},
  \bibinfo{person}{Edward~G. Hohenstein}, \bibinfo{person}{Uğur Bozkaya},
  \bibinfo{person}{Alexander~Yu. Sokolov}, \bibinfo{person}{Roberto
  Di~Remigio}, \bibinfo{person}{Ryan~M. Richard}, \bibinfo{person}{Jérôme~F.
  Gonthier}, \bibinfo{person}{Andrew~M. James}, \bibinfo{person}{Harley~R.
  McAlexander}, \bibinfo{person}{Ashutosh Kumar}, \bibinfo{person}{Masaaki
  Saitow}, \bibinfo{person}{Xiao Wang}, \bibinfo{person}{Benjamin~P.
  Pritchard}, \bibinfo{person}{Prakash Verma}, \bibinfo{person}{Henry F.~III
  Schaefer}, \bibinfo{person}{Konrad Patkowski}, \bibinfo{person}{Rollin~A.
  King}, \bibinfo{person}{Edward~F. Valeev}, \bibinfo{person}{Francesco~A.
  Evangelista}, \bibinfo{person}{Justin~M. Turney}, \bibinfo{person}{T.~Daniel
  Crawford}, {and} \bibinfo{person}{C.~David Sherrill}.}
  \bibinfo{year}{2017}\natexlab{}.
\newblock \showarticletitle{Psi4 1.1: {An} {Open}-{Source} {Electronic}
  {Structure} {Program} {Emphasizing} {Automation}, {Advanced} {Libraries}, and
  {Interoperability}}.
\newblock \bibinfo{journal}{\emph{J. Chem. Theory Comput.}}
  \bibinfo{volume}{13}, \bibinfo{number}{7} (\bibinfo{date}{July}
  \bibinfo{year}{2017}), \bibinfo{pages}{3185--3197}.
\newblock
\showISSN{1549-9618}
\href{https://doi.org/10.1021/acs.jctc.7b00174}{doi:\nolinkurl{10.1021/acs.jctc.7b00174}}


\bibitem[Patel et~al\mbox{.}(2024)]%
        {Patel2024-qv}
\bibfield{author}{\bibinfo{person}{Yash~J Patel}, \bibinfo{person}{Akash
  Kundu}, \bibinfo{person}{Mateusz Ostaszewski}, \bibinfo{person}{Xavier
  Bonet-Monroig}, \bibinfo{person}{Vedran Dunjko}, {and} \bibinfo{person}{Onur
  Danaci}.} \bibinfo{year}{2024}\natexlab{}.
\newblock \bibinfo{title}{Curriculum reinforcement learning for quantum
  architecture search under hardware errors}.
\newblock


\bibitem[Pedregosa et~al\mbox{.}(2011a)]%
        {pedregosa_scikit-learn_2011}
\bibfield{author}{\bibinfo{person}{F. Pedregosa}, \bibinfo{person}{G.
  Varoquaux}, \bibinfo{person}{A. Gramfort}, \bibinfo{person}{V. Michel},
  \bibinfo{person}{B. Thirion}, \bibinfo{person}{O. Grisel},
  \bibinfo{person}{M. Blondel}, \bibinfo{person}{P. Prettenhofer},
  \bibinfo{person}{R. Weiss}, \bibinfo{person}{V. Dubourg}, \bibinfo{person}{J.
  Vanderplas}, \bibinfo{person}{A. Passos}, \bibinfo{person}{D. Cournapeau},
  \bibinfo{person}{M. Brucher}, \bibinfo{person}{M. Perrot}, {and}
  \bibinfo{person}{E. Duchesnay}.} \bibinfo{year}{2011}\natexlab{a}.
\newblock \showarticletitle{Scikit-learn: {Machine} {Learning} in {Python}}.
\newblock \bibinfo{journal}{\emph{J. Mach. Learn. Res.}}  \bibinfo{volume}{12}
  (\bibinfo{year}{2011}), \bibinfo{pages}{2825--2830}.
\newblock


\bibitem[Pedregosa et~al\mbox{.}(2011b)]%
        {Pedregosa2011-bf}
\bibfield{author}{\bibinfo{person}{F Pedregosa}, \bibinfo{person}{G Varoquaux},
  \bibinfo{person}{A Gramfort}, \bibinfo{person}{V Michel}, \bibinfo{person}{B
  Thirion}, \bibinfo{person}{O Grisel}, \bibinfo{person}{M Blondel},
  \bibinfo{person}{P Prettenhofer}, \bibinfo{person}{R Weiss},
  \bibinfo{person}{V Dubourg}, \bibinfo{person}{J Vanderplas},
  \bibinfo{person}{A Passos}, \bibinfo{person}{D Cournapeau},
  \bibinfo{person}{M Brucher}, \bibinfo{person}{M Perrot}, {and}
  \bibinfo{person}{E Duchesnay}.} \bibinfo{year}{2011}\natexlab{b}.
\newblock \showarticletitle{Scikit-learn: Machine Learning in {P}ython}.
\newblock \bibinfo{journal}{\emph{J. Mach. Learn. Res.}}  \bibinfo{volume}{12}
  (\bibinfo{year}{2011}), \bibinfo{pages}{2825--2830}.
\newblock


\bibitem[P{\'e}rez-Salinas et~al\mbox{.}(2020)]%
        {perez2020data}
\bibfield{author}{\bibinfo{person}{Adri{\'a}n P{\'e}rez-Salinas},
  \bibinfo{person}{Alba Cervera-Lierta}, \bibinfo{person}{Elies Gil-Fuster},
  {and} \bibinfo{person}{Jos{\'e}~I Latorre}.} \bibinfo{year}{2020}\natexlab{}.
\newblock \showarticletitle{Data re-uploading for a universal quantum
  classifier}.
\newblock \bibinfo{journal}{\emph{Quantum}}  \bibinfo{volume}{4}
  (\bibinfo{year}{2020}), \bibinfo{pages}{226}.
\newblock


\bibitem[Peruzzo et~al\mbox{.}(2014)]%
        {Peruzzo2014-gh}
\bibfield{author}{\bibinfo{person}{Alberto Peruzzo}, \bibinfo{person}{Jarrod
  McClean}, \bibinfo{person}{Peter Shadbolt}, \bibinfo{person}{Man-Hong Yung},
  \bibinfo{person}{Xiao-Qi Zhou}, \bibinfo{person}{Peter~J Love},
  \bibinfo{person}{Al{\'a}n Aspuru-Guzik}, {and} \bibinfo{person}{Jeremy~L
  O’brien}.} \bibinfo{year}{2014}\natexlab{}.
\newblock \showarticletitle{A variational eigenvalue solver on a photonic
  quantum processor}.
\newblock \bibinfo{journal}{\emph{Nat. Commun.}} \bibinfo{volume}{5},
  \bibinfo{number}{1} (\bibinfo{year}{2014}), \bibinfo{pages}{4213}.
\newblock


\bibitem[Prasad et~al\mbox{.}(2021)]%
        {Prasad2021-sx}
\bibfield{author}{\bibinfo{person}{Viki~Kumar Prasad},
  \bibinfo{person}{M~Hossein Khalilian}, \bibinfo{person}{Alberto Otero-de-la
  Roza}, {and} \bibinfo{person}{Gino~A DiLabio}.}
  \bibinfo{year}{2021}\natexlab{}.
\newblock \showarticletitle{{BSE49}, a diverse, high-quality benchmark dataset
  of separation energies of chemical bonds}.
\newblock \bibinfo{journal}{\emph{Sci. Data}} \bibinfo{volume}{8},
  \bibinfo{number}{1} (\bibinfo{year}{2021}), \bibinfo{pages}{300}.
\newblock


\bibitem[Raccuglia et~al\mbox{.}(2016)]%
        {Raccuglia2016-tw}
\bibfield{author}{\bibinfo{person}{Paul Raccuglia},
  \bibinfo{person}{Katherine~C Elbert}, \bibinfo{person}{Philip D~F Adler},
  \bibinfo{person}{Casey Falk}, \bibinfo{person}{Malia~B Wenny},
  \bibinfo{person}{Aurelio Mollo}, \bibinfo{person}{Matthias Zeller},
  \bibinfo{person}{Sorelle~A Friedler}, \bibinfo{person}{Joshua Schrier}, {and}
  \bibinfo{person}{Alexander~J Norquist}.} \bibinfo{year}{2016}\natexlab{}.
\newblock \showarticletitle{Machine-learning-assisted materials discovery using
  failed experiments}.
\newblock \bibinfo{journal}{\emph{Nature}} \bibinfo{volume}{533},
  \bibinfo{number}{7601} (\bibinfo{date}{May} \bibinfo{year}{2016}),
  \bibinfo{pages}{73--76}.
\newblock


\bibitem[Ramakrishnan et~al\mbox{.}(2014)]%
        {Ramakrishnan2014-oj}
\bibfield{author}{\bibinfo{person}{Raghunathan Ramakrishnan},
  \bibinfo{person}{Pavlo~O Dral}, \bibinfo{person}{Matthias Rupp}, {and}
  \bibinfo{person}{O~Anatole von Lilienfeld}.} \bibinfo{year}{2014}\natexlab{}.
\newblock \showarticletitle{Quantum chemistry structures and properties of 134
  kilo molecules}.
\newblock \bibinfo{journal}{\emph{Sci. Data}} \bibinfo{volume}{1},
  \bibinfo{number}{1} (\bibinfo{date}{Aug.} \bibinfo{year}{2014}),
  \bibinfo{pages}{140022}.
\newblock


\bibitem[Ramakrishnan et~al\mbox{.}(2015)]%
        {Ramakrishnan2015-dg}
\bibfield{author}{\bibinfo{person}{Raghunathan Ramakrishnan},
  \bibinfo{person}{Pavlo~O Dral}, \bibinfo{person}{Matthias Rupp}, {and}
  \bibinfo{person}{O~Anatole Von~Lilienfeld}.} \bibinfo{year}{2015}\natexlab{}.
\newblock \showarticletitle{Big data meets quantum chemistry approximations:
  the {$\Delta$}-machine learning approach}.
\newblock \bibinfo{journal}{\emph{J. Chem. Theory Comput.}}
  \bibinfo{volume}{11}, \bibinfo{number}{5} (\bibinfo{year}{2015}),
  \bibinfo{pages}{2087--2096}.
\newblock


\bibitem[Rogers and Hahn(2010)]%
        {Rogers2010-yp}
\bibfield{author}{\bibinfo{person}{David Rogers} {and} \bibinfo{person}{Mathew
  Hahn}.} \bibinfo{year}{2010}\natexlab{}.
\newblock \showarticletitle{Extended-connectivity fingerprints}.
\newblock \bibinfo{journal}{\emph{J. Chem. Inf. Model.}} \bibinfo{volume}{50},
  \bibinfo{number}{5} (\bibinfo{date}{May} \bibinfo{year}{2010}),
  \bibinfo{pages}{742--754}.
\newblock


\bibitem[Sanchez-Lengeling and Aspuru-Guzik(2018)]%
        {Sanchez-Lengeling2018-xt}
\bibfield{author}{\bibinfo{person}{Benjamin Sanchez-Lengeling} {and}
  \bibinfo{person}{Alán Aspuru-Guzik}.} \bibinfo{year}{2018}\natexlab{}.
\newblock \showarticletitle{Inverse molecular design using machine learning:
  Generative models for matter engineering}.
\newblock \bibinfo{journal}{\emph{Science}} \bibinfo{volume}{361},
  \bibinfo{number}{6400} (\bibinfo{date}{July} \bibinfo{year}{2018}),
  \bibinfo{pages}{360--365}.
\newblock


\bibitem[Schuld(2021)]%
        {Schuld2021-gb}
\bibfield{author}{\bibinfo{person}{Maria Schuld}.}
  \bibinfo{year}{2021}\natexlab{}.
\newblock \showarticletitle{Supervised quantum machine learning models are
  kernel methods}.
\newblock \bibinfo{journal}{\emph{arXiv preprint arXiv:2101.11020}}
  (\bibinfo{year}{2021}).
\newblock


\bibitem[Schuld et~al\mbox{.}(2015)]%
        {schuld2015introduction}
\bibfield{author}{\bibinfo{person}{Maria Schuld}, \bibinfo{person}{Ilya
  Sinayskiy}, {and} \bibinfo{person}{Francesco Petruccione}.}
  \bibinfo{year}{2015}\natexlab{}.
\newblock \showarticletitle{An introduction to quantum machine learning}.
\newblock \bibinfo{journal}{\emph{Contemp. Phys.}} \bibinfo{volume}{56},
  \bibinfo{number}{2} (\bibinfo{year}{2015}), \bibinfo{pages}{172--185}.
\newblock


\bibitem[Schuld et~al\mbox{.}(2021)]%
        {Schuld2021-sf}
\bibfield{author}{\bibinfo{person}{Maria Schuld}, \bibinfo{person}{Ryan Sweke},
  {and} \bibinfo{person}{Johannes~Jakob Meyer}.}
  \bibinfo{year}{2021}\natexlab{}.
\newblock \showarticletitle{Effect of data encoding on the expressive power of
  variational quantum-machine-learning models}.
\newblock \bibinfo{journal}{\emph{Phys. Rev. A}}  \bibinfo{volume}{103}
  (\bibinfo{date}{March} \bibinfo{year}{2021}), \bibinfo{pages}{032430}.
\newblock


\bibitem[Smith et~al\mbox{.}(2018)]%
        {smith_psi4numpy_2018}
\bibfield{author}{\bibinfo{person}{Daniel G.~A. Smith},
  \bibinfo{person}{Lori~A. Burns}, \bibinfo{person}{Dominic~A. Sirianni},
  \bibinfo{person}{Daniel~R. Nascimento}, \bibinfo{person}{Ashutosh Kumar},
  \bibinfo{person}{Andrew~M. James}, \bibinfo{person}{Jeffrey~B. Schriber},
  \bibinfo{person}{Tianyuan Zhang}, \bibinfo{person}{Boyi Zhang},
  \bibinfo{person}{Adam~S. Abbott}, \bibinfo{person}{Eric~J. Berquist},
  \bibinfo{person}{Marvin~H. Lechner}, \bibinfo{person}{Leonardo~A. Cunha},
  \bibinfo{person}{Alexander~G. Heide}, \bibinfo{person}{Jonathan~M. Waldrop},
  \bibinfo{person}{Tyler~Y. Takeshita}, \bibinfo{person}{Asem Alenaizan},
  \bibinfo{person}{Daniel Neuhauser}, \bibinfo{person}{Rollin~A. King},
  \bibinfo{person}{Andrew~C. Simmonett}, \bibinfo{person}{Justin~M. Turney},
  \bibinfo{person}{Henry~F. Schaefer}, \bibinfo{person}{Francesco~A.
  Evangelista}, \bibinfo{person}{A.~Eugene~III DePrince},
  \bibinfo{person}{T.~Daniel Crawford}, \bibinfo{person}{Konrad Patkowski},
  {and} \bibinfo{person}{C.~David Sherrill}.} \bibinfo{year}{2018}\natexlab{}.
\newblock \showarticletitle{{Psi4NumPy}: {An} {Interactive} {Quantum}
  {Chemistry} {Programming} {Environment} for {Reference} {Implementations} and
  {Rapid} {Development}}.
\newblock \bibinfo{journal}{\emph{J. Chem. Theory Comput.}}
  \bibinfo{volume}{14}, \bibinfo{number}{7} (\bibinfo{date}{July}
  \bibinfo{year}{2018}), \bibinfo{pages}{3504--3511}.
\newblock
\showISSN{1549-9618}
\href{https://doi.org/10.1021/acs.jctc.8b00286}{doi:\nolinkurl{10.1021/acs.jctc.8b00286}}


\bibitem[Stanley and Miikkulainen(2002)]%
        {Stanley2002-rj}
\bibfield{author}{\bibinfo{person}{Kenneth~O Stanley} {and}
  \bibinfo{person}{Risto Miikkulainen}.} \bibinfo{year}{2002}\natexlab{}.
\newblock \showarticletitle{Evolving Neural Networks through Augmenting
  Topologies}.
\newblock \bibinfo{journal}{\emph{Evol. Comput.}} \bibinfo{volume}{10},
  \bibinfo{number}{2} (\bibinfo{year}{2002}), \bibinfo{pages}{99--127}.
\newblock


\bibitem[Suzuki and Katouda(2020)]%
        {Suzuki2020-gv}
\bibfield{author}{\bibinfo{person}{Teppei Suzuki} {and} \bibinfo{person}{Michio
  Katouda}.} \bibinfo{year}{2020}\natexlab{}.
\newblock \showarticletitle{Predicting toxicity by quantum machine learning}.
\newblock \bibinfo{journal}{\emph{J. Phys. Commun.}} \bibinfo{volume}{4},
  \bibinfo{number}{12} (\bibinfo{date}{Dec.} \bibinfo{year}{2020}),
  \bibinfo{pages}{125012}.
\newblock


\bibitem[Suzuki et~al\mbox{.}(2021)]%
        {Suzuki2020-pc}
\bibfield{author}{\bibinfo{person}{Yasunari Suzuki}, \bibinfo{person}{Yoshiaki
  Kawase}, \bibinfo{person}{Yuya Masumura}, \bibinfo{person}{Yuria Hiraga},
  \bibinfo{person}{Masahiro Nakadai}, \bibinfo{person}{Jiabao Chen},
  \bibinfo{person}{Ken~M Nakanishi}, \bibinfo{person}{Kosuke Mitarai},
  \bibinfo{person}{Ryosuke Imai}, \bibinfo{person}{Shiro Tamiya},
  {et~al\mbox{.}}} \bibinfo{year}{2021}\natexlab{}.
\newblock \showarticletitle{Qulacs: a fast and versatile quantum circuit
  simulator for research purpose}.
\newblock \bibinfo{journal}{\emph{Quantum}}  \bibinfo{volume}{5}
  (\bibinfo{year}{2021}), \bibinfo{pages}{559}.
\newblock


\bibitem[Townsend and Vogiatzis(2019)]%
        {Townsend2019-gx}
\bibfield{author}{\bibinfo{person}{Jacob Townsend} {and}
  \bibinfo{person}{Konstantinos~D Vogiatzis}.} \bibinfo{year}{2019}\natexlab{}.
\newblock \showarticletitle{Data-Driven Acceleration of the Coupled-Cluster
  Singles and Doubles Iterative Solver}.
\newblock \bibinfo{journal}{\emph{J. Phys. Chem. Lett.}} \bibinfo{volume}{10},
  \bibinfo{number}{14} (\bibinfo{date}{June} \bibinfo{year}{2019}),
  \bibinfo{pages}{4129–4135}.
\newblock


\bibitem[Unke and Meuwly(2019)]%
        {Unke2019-xz}
\bibfield{author}{\bibinfo{person}{Oliver~T Unke} {and} \bibinfo{person}{Markus
  Meuwly}.} \bibinfo{year}{2019}\natexlab{}.
\newblock \showarticletitle{PhysNet: A neural network for predicting energies,
  forces, dipole moments, and partial charges}.
\newblock \bibinfo{journal}{\emph{J. Chem. Theory Comput.}}
  \bibinfo{volume}{15}, \bibinfo{number}{6} (\bibinfo{year}{2019}),
  \bibinfo{pages}{3678--3693}.
\newblock


\bibitem[van~den Berg et~al\mbox{.}(2022)]%
        {van_den_berg_model-free_2022}
\bibfield{author}{\bibinfo{person}{Ewout van~den Berg},
  \bibinfo{person}{Zlatko~K. Minev}, {and} \bibinfo{person}{Kristan Temme}.}
  \bibinfo{year}{2022}\natexlab{}.
\newblock \showarticletitle{Model-free readout-error mitigation for quantum
  expectation values}.
\newblock \bibinfo{journal}{\emph{Phys. Rev. A}} \bibinfo{volume}{105},
  \bibinfo{number}{3} (\bibinfo{date}{March} \bibinfo{year}{2022}),
  \bibinfo{pages}{032620}.
\newblock
\href{https://doi.org/10.1103/PhysRevA.105.032620}{doi:\nolinkurl{10.1103/PhysRevA.105.032620}}


\bibitem[Virtanen et~al\mbox{.}(2020)]%
        {virtanen_scipy_2020}
\bibfield{author}{\bibinfo{person}{Pauli Virtanen}, \bibinfo{person}{Ralf
  Gommers}, \bibinfo{person}{Travis~E Oliphant}, \bibinfo{person}{Matt
  Haberland}, \bibinfo{person}{Tyler Reddy}, \bibinfo{person}{David
  Cournapeau}, \bibinfo{person}{Evgeni Burovski}, \bibinfo{person}{Pearu
  Peterson}, \bibinfo{person}{Warren Weckesser}, \bibinfo{person}{Jonathan
  Bright}, {et~al\mbox{.}}} \bibinfo{year}{2020}\natexlab{}.
\newblock \showarticletitle{SciPy 1.0: fundamental algorithms for scientific
  computing in Python}.
\newblock \bibinfo{journal}{\emph{Nat. Methods}} \bibinfo{volume}{17},
  \bibinfo{number}{3} (\bibinfo{year}{2020}), \bibinfo{pages}{261--272}.
\newblock


\bibitem[Wang et~al\mbox{.}(2022)]%
        {Wang2022-hi}
\bibfield{author}{\bibinfo{person}{Hanrui Wang}, \bibinfo{person}{Yongshan
  Ding}, \bibinfo{person}{Jiaqi Gu}, \bibinfo{person}{Yujun Lin},
  \bibinfo{person}{David~Z Pan}, \bibinfo{person}{Frederic~T Chong}, {and}
  \bibinfo{person}{Song Han}.} \bibinfo{year}{2022}\natexlab{}.
\newblock \showarticletitle{{QuantumNAS}: Noise-Adaptive Search for Robust
  Quantum Circuits}. In \bibinfo{booktitle}{\emph{2022 IEEE International
  Symposium on High-Performance Computer Architecture (HPCA)}}.
  \bibinfo{publisher}{IEEE}, \bibinfo{pages}{692–708}.
\newblock


\bibitem[Wood et~al\mbox{.}(2006)]%
        {Wood2006-zn}
\bibfield{author}{\bibinfo{person}{Geoffrey P~F Wood}, \bibinfo{person}{Leo
  Radom}, \bibinfo{person}{George~A Petersson}, \bibinfo{person}{Ericka~C
  Barnes}, \bibinfo{person}{Michael~J Frisch}, {and} \bibinfo{person}{John~A
  Montgomery, Jr}.} \bibinfo{year}{2006}\natexlab{}.
\newblock \showarticletitle{A restricted-open-shell complete-basis-set model
  chemistry}.
\newblock \bibinfo{journal}{\emph{J. Chem. Phys.}} \bibinfo{volume}{125},
  \bibinfo{number}{9} (\bibinfo{date}{Sept.} \bibinfo{year}{2006}),
  \bibinfo{pages}{094106}.
\newblock


\bibitem[Yang et~al\mbox{.}(2019a)]%
        {Yang2019-fx}
\bibfield{author}{\bibinfo{person}{Kevin Yang}, \bibinfo{person}{Kyle Swanson},
  \bibinfo{person}{Wengong Jin}, \bibinfo{person}{Connor Coley},
  \bibinfo{person}{Philipp Eiden}, \bibinfo{person}{Hua Gao},
  \bibinfo{person}{Angel Guzman-Perez}, \bibinfo{person}{Timothy Hopper},
  \bibinfo{person}{Brian Kelley}, \bibinfo{person}{Miriam Mathea},
  {et~al\mbox{.}}} \bibinfo{year}{2019}\natexlab{a}.
\newblock \showarticletitle{Analyzing learned molecular representations for
  property prediction}.
\newblock \bibinfo{journal}{\emph{J. Chem. Inf. Model.}} \bibinfo{volume}{59},
  \bibinfo{number}{8} (\bibinfo{year}{2019}), \bibinfo{pages}{3370--3388}.
\newblock


\bibitem[Yang et~al\mbox{.}(2019b)]%
        {Yang2019-jj}
\bibfield{author}{\bibinfo{person}{Xin Yang}, \bibinfo{person}{Yifei Wang},
  \bibinfo{person}{Ryan Byrne}, \bibinfo{person}{Gisbert Schneider}, {and}
  \bibinfo{person}{Shengyong Yang}.} \bibinfo{year}{2019}\natexlab{b}.
\newblock \showarticletitle{Concepts of artificial intelligence for
  computer-assisted drug discovery}.
\newblock \bibinfo{journal}{\emph{Chem. Rev.}} \bibinfo{volume}{119},
  \bibinfo{number}{18} (\bibinfo{date}{Sept.} \bibinfo{year}{2019}),
  \bibinfo{pages}{10520--10594}.
\newblock


\bibitem[Zhang et~al\mbox{.}(2022)]%
        {Zhang2022-ho}
\bibfield{author}{\bibinfo{person}{Shi-Xin Zhang}, \bibinfo{person}{Chang-Yu
  Hsieh}, \bibinfo{person}{Shengyu Zhang}, {and} \bibinfo{person}{Hong Yao}.}
  \bibinfo{year}{2022}\natexlab{}.
\newblock \showarticletitle{Differentiable quantum architecture search}.
\newblock \bibinfo{journal}{\emph{Quantum Sci. Technol.}} \bibinfo{volume}{7},
  \bibinfo{number}{4} (\bibinfo{date}{Aug.} \bibinfo{year}{2022}),
  \bibinfo{pages}{045023}.
\newblock


\bibitem[Zhong et~al\mbox{.}(2020)]%
        {Zhong2020-dx}
\bibfield{author}{\bibinfo{person}{Miao Zhong}, \bibinfo{person}{Kevin Tran},
  \bibinfo{person}{Yimeng Min}, \bibinfo{person}{Chuanhao Wang},
  \bibinfo{person}{Ziyun Wang}, \bibinfo{person}{Cao-Thang Dinh},
  \bibinfo{person}{Phil De~Luna}, \bibinfo{person}{Zongqian Yu},
  \bibinfo{person}{Armin~Sedighian Rasouli}, \bibinfo{person}{Peter Brodersen},
  \bibinfo{person}{Song Sun}, \bibinfo{person}{Oleksandr Voznyy},
  \bibinfo{person}{Chih-Shan Tan}, \bibinfo{person}{Mikhail Askerka},
  \bibinfo{person}{Fanglin Che}, \bibinfo{person}{Min Liu},
  \bibinfo{person}{Ali Seifitokaldani}, \bibinfo{person}{Yuanjie Pang},
  \bibinfo{person}{Shen-Chuan Lo}, \bibinfo{person}{Alexander Ip},
  \bibinfo{person}{Zachary Ulissi}, {and} \bibinfo{person}{Edward~H Sargent}.}
  \bibinfo{year}{2020}\natexlab{}.
\newblock \showarticletitle{Accelerated discovery of {CO2} electrocatalysts
  using active machine learning}.
\newblock \bibinfo{journal}{\emph{Nature}} \bibinfo{volume}{581},
  \bibinfo{number}{7807} (\bibinfo{date}{May} \bibinfo{year}{2020}),
  \bibinfo{pages}{178--183}.
\newblock


\end{thebibliography}

\end{document}